\documentclass[journal=jacsat,manuscript=article]{achemso}

\usepackage[version=3]{mhchem} 
\usepackage{titlesec}
\titlelabel{\thetitle.\quad}
\usepackage{subcaption}
\usepackage{graphicx}
\usepackage{subcaption}
\usepackage[table,xcdraw]{xcolor}
\usepackage[table,xcdraw]{xcolor}
\usepackage[table,xcdraw]{xcolor}
\usepackage{tabularx}
\usepackage{soul}
\sethlcolor{yellow}

\author{Mahya Ghorab}
\affiliation[Constructor University]
{School of Computer Science \& Engineering, Constructor University, Bremen, Germany}
\author{Ayush K. Ranga}
\affiliation[Constructor University]
{School of Science, Constructor University, Bremen, Germany}
\author{Patrice Donfack}
\affiliation[Constructor University]
{School of Science, Constructor University, Bremen, Germany}
\author{Arnulf Materny}
\affiliation[Constructor University]
{School of Science, Constructor University, Bremen, Germany}
\author{Veit Wagner}
\affiliation[Constructor University]
{School of Science, Constructor University, Bremen, Germany}
\author{Mojtaba Joodaki}
\email{mjoodaki@constructor.university}
\phone{+49 421 200-3215}
\fax{+49 421 200-3103}
\affiliation[Constructor University]
{School of Computer Science \& Engineering, Constructor University, Bremen, Germany}

\title[An \textsf{achemso} demo]
  {Strain-Induced Optical and Molecular Transformations in PET Films for Organic Electronic Applications}

\abbreviations{IR,NMR,UV}
\keywords{American Chemical Society, \LaTeX}

\begin{document}

\begin{tocentry}

Some journals require a graphical entry for the Table of Contents.
This should be laid out ``print ready'' so that the sizing of the
text is correct.

Inside the \texttt{tocentry} environment, the font used is Helvetica
8\,pt, as required by \emph{Journal of the American Chemical
Society}.

The surrounding frame is 9\,cm by 3.5\,cm, which is the maximum
permitted for  \emph{Journal of the American Chemical Society}
graphical table of content entries. The box will not resize if the
content is too big: instead it will overflow the edge of the box.

This box and the associated title will always be printed on a
separate page at the end of the document.

\end{tocentry}

\begin{abstract}
 Poly(ethylene terephthalate) (PET) films are widely used in flexible electronics and optoelectronics, where their mechanical durability and optical performance under strain are essential for device reliability. This study investigates the impact of applied mechanical strain on the optical and molecular properties of PET at room temperature, using UV-Vis absorption and Raman spectroscopy. The work explores how varying strain levels, from 0\% (unstretched) to 30\%, affect the transparency, vibrational modes, and molecular reorganization within PET films. UV-Vis absorbance measurements reveal that strain induces significant changes in the light transmission properties of PET, particularly in the visible range, and increases absorption in the UVA and visible region by up to 100\%. Raman spectra indicate that strain levels higher than 5\% lead to irreversible shifts of vibrational lines, accompanied by an increase of their full width at half maximum (FWHM), suggesting molecular reorientation and crystallinity changes. The phonon mode coupled with C-O stretching [O-CH$_{2}$] shows the strongest response to applied mechanical stress. This study provides a comprehensive understanding of strain-induced optical and structural alterations in PET, with implications for improving the mechanical and optical performance of PET-based devices in strain-sensitive applications, such as organic solar cells (OSCs), organic light-emitting diodes (OLEDs), and flexible sensors.
\end{abstract}

\section{Introduction}
Flexible and stretchable materials are the driving force of innovation in technologies such as organic solar cells (OSCs), organic light-emitting diodes (OLEDs), and strain sensors, where mechanical durability and optical performance under strain are critical for reliable operation\cite{MPaper, Q.Jing, ghorab2024}. Poly(ethylene terephthalate) (PET) is a well-known material in such applications due to its optical properties, mechanical robustness, and ease of processing \cite{Wagner, Oksana}. Despite PET’s widespread applications, a detailed understanding of how mechanical strain influences its molecular structure and optical behavior remains incomplete. This knowledge is crucial for optimizing PET-based device architectures and ensuring stability under operational stresses \cite{Gupta, Bin, Lim}.

PET is the primary interface interacting with light in OSCs before reaching the energy-converting active layer. Hence, its optical properties, particularly in the visible range, are critical for maximizing light transmission to underlying layers \cite{Dauzon, Wageh}. 

Mechanical strain, such as stretching or compression, induces structural changes in PET, including molecular reorientation and variations in polymer crystallinity due to chain alignment. These alterations affect material transparency by influencing light scattering, absorption, and refractive index. Investigating these changes helps evaluate PET's degradation under mechanical stress and its impact on optical properties, ensuring reliability in flexible electronic applications \cite{Teng,JOODAKI2018230}.

Previously, studies have investigated the impact of strain and external conditions on the molecular structure and properties of PET \cite{TONG, Celik2022}. Webster and Bower demonstrated a linear relation between hydrostatic pressure and the Raman shift of the 1616 cm$^{-1}$ mode, showing how pressure alters intermolecular interactions and molecular packing. This work set a foundational understanding of how PET responds to isotropic stress \cite{Webster}. Bin \textit{et al.} examined the effects of biaxial stretching on PET, revealing changes in molecular orientation and crystallinity, which strongly influence its mechanical properties, such as Young’s modulus, and optical features, such as birefringence \cite{Bin}. Similarly, Laskarakis \textit{et al.} explored the electronic and vibrational properties of PET under biaxial stretching during industrial processing. This study emphasized the impact of strain on the optical anisotropy of PET, particularly through changes in electronic transitions as well as changes in the dielectric function across the visible to far-ultraviolet spectrum. These findings complement the mechanical and molecular alignment studies by Bin \textit{et al.}, providing a better understanding of how strain alters the electronic and vibrational behavior of PET \cite{Laskarakis}. González-Córdova \textit{et al.} utilized angular-dependent Raman scattering to study strain-induced anisotropy in PET under uniaxial strain, highlighting significant changes in molecular symmetry and the polarization dependence of Raman intensities \cite{GonzalezCordova}.

However, most of these studies were conducted under elevated-temperature conditions to facilitate molecular mobility and alignment during biaxial stretching experiments. For example, the biaxial stretching experiments by Bin \textit{et al.} and Laskarakis \textit{et al.} took place at temperatures above the PET glass transition point ($\sim$80--90$^\circ$C), where molecular chains gain sufficient mobility to undergo significant rearrangements. In contrast, the current study investigates the response of PET to strain at room temperature, where restricted molecular mobility leads to phenomena different from those observed at elevated temperatures. As discussed later in the paper, the room-temperature conditions result in reduced crystallinity evolution and a more pronounced influence of the amorphous regions, giving rise to a unique vibrational and optical behavior not previously reported.

Furthermore, earlier studies often focused on narrow strain ranges, shorter timeframes, or a limited number of vibrational modes, leaving gaps in understanding PET’s molecular deformation mechanisms. This study addresses these limitations by comprehensively investigating a broad strain range (1–30\%) over extended periods and analyzing 13 Raman modes. It correlates frequency shifts, full-width half-maximum (FWHM) variations, and intensity changes, bridging existing gaps to provide a holistic understanding of strain-induced molecular and optical changes. These findings offer critical insights for flexible electronics, sensors, and optical device applications.

Previous studies, such as the investigation by Webster \textit{et al.} \cite{Webster}, have demonstrated that Raman signals taken with the polarization of the exciting laser parallel to the strain axis exhibit a significantly more pronounced response to the stretching than those taken with perpendicular polarization. This increased sensitivity allows a better understanding of molecular reorganization and vibration mode responses under mechanical deformation. Moreover, fixed polarization simulates light interactions in anisotropic PET substrates, influencing optical transmission during device fabrication and operation.

This work has captured the dominant strain-induced molecular interactions by focusing on the parallel excitation polarization, providing a deeper understanding of the vibration modes responsive to strain. By addressing points that have not yet been fully understood in earlier studies for these vibration modes and linking molecular reorganization to optical property changes, this study improves the understanding of PET's mechanical and optical resilience by offering pathways to design strain-sensitive devices and improve PET-based device architectures.

\section{Experimental Methods and Setup}

This work has utilized UV-Vis absorption and Raman spectroscopy to examine the optical and vibrational responses of poly(ethylene terephthalate) (PET) films under mechanical strain. 

\begin{figure}[h]
    \centering
        \includegraphics[width=\textwidth]{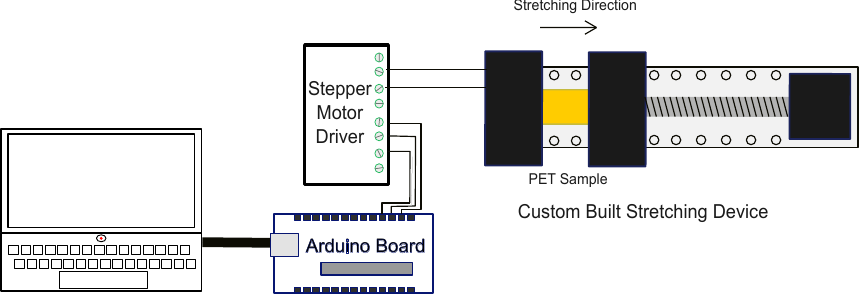}
 \caption{ Custom-built stretching device; the stretcher is controlled by an Arduino board, via a motor controller.}
    \label{Experiment setup}
\end {figure}

Figure \ref{Experiment setup} illustrates the custom-built stretcher device designed to apply precise, controlled tensile strain to PET samples. Measurements were performed both in-situ and ex-situ, depending on the experimental technique. The device is controlled by an Arduino board, which regulates the application of strain. PET films were securely clamped at both ends of the stretcher and incrementally strained at levels of 1\%, 2\%, 3\%, 4\%, 5\%, 10\%, 15\%, 20\%, 25\%, and 30\%.

UV-Vis absorption spectroscopy offers a straightforward, fast, and cost-effective method to assess the absorbance and transmittance of materials in the ultraviolet and visible wavelength ranges. This technique is particularly valuable for quantifying changes in transparency resulting from mechanical strain \cite{Oconnor, Akash2020}. For PET films under strain, UV-Vis absorption measurements capture variations in absorbance across specific wavelength ranges, providing a rapid and quantitative evaluation of transparency. Such changes in optical behavior often reflect underlying molecular electronic or structural alterations. However, while UV-Vis absorption spectroscopy effectively identifies changes in transparency, it offers limited insight into the molecular mechanisms driving these changes.

In contrast, Raman spectroscopy provides a deeper understanding of structural dynamics by probing molecular vibrations directly linked to the material’s molecular architecture. Raman spectroscopy enables in-situ measurements as a non-destructive technique, allowing real-time observation of structural variations under strain \cite{Kuznetsov2021, Dietzek2018}. By correlating Raman data with UV-Vis findings, it becomes possible to detect these transparency changes and elucidate their causes by identifying the specific molecular mechanisms responsible. This synergistic approach combines the broad applicability of UV-Vis with the structural precision of Raman, offering a comprehensive understanding of PET’s response to mechanical stress. 

UV-Vis absorbance measurements were performed on the PET film after each strain exposure. The PET roll used in this study was sourced industrially, with a density of $16.8 \times 10^5$ $\text{g/m}^3$. The films were cut into standardized dimensions of 15 $\text{mm}$ $\times$ 33 $\text{mm}$ to fit the UV-Vis device sample holder size and ensure uniformity throughout the experiments. Each strain level was maintained for 30 minutes for structural relaxation and accurate spectral measurements. A new, unstretched film was used for each strain value to ensure consistent and independent measurements. The strain was calculated as the percentage elongation relative to the original film length ($\epsilon = \frac{\Delta L}{L}$).

\begin{figure}[]
    \centering
\includegraphics[width=0.5\textwidth]{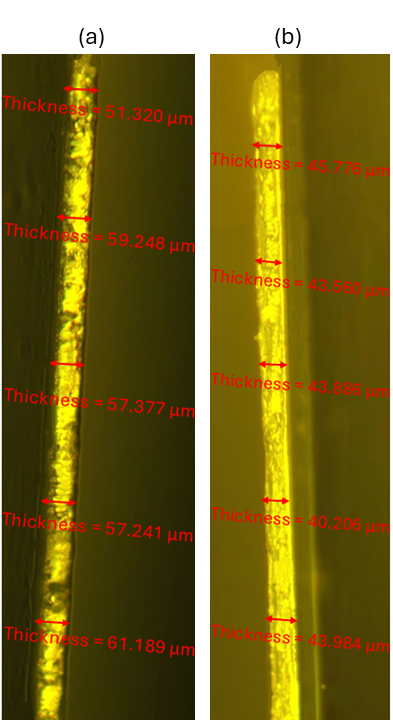}
   \caption{The PET sample's thickness was measured at different positions. (a) without stretching and (b) after 20\% stretching; the thickness deduction resulting from the strain exposure is observed. The images were captured using a Carl Zeiss Axiotech 100HD light microscope equipped with an Axiocam 105 color camera.}
   \label{cross-section}
\end {figure}

 While UV-Vis absorption spectroscopy under strain (in-situ) is undoubtedly valuable for capturing real-time changes in optical and electronic properties, in our case, the experimental setup posed a limitation. The stretcher device could not be accommodated within our UV-Vis spectrometer, making in-situ (i.e., in the stretched state) measurements infeasible. As a result, we focused on ex-situ UV-Vis absorption measurements, which also resulted in valuable data. This approach allows us to evaluate the optical properties of the material after the application and release of strain, offering crucial information about the reversibility of changes, the presence of residual stress, and potential permanent structural modifications. Despite the limitation, the ex-situ measurements provided a robust understanding of the effects of strain on the material, highlighting the significance of post-strain optical analysis in assessing material stability and mechanical behavior. This test was performed using a Perkin Elmer Lambda 12 spectrophotometer. The PET samples were directly mounted in the beam path without using a cuvette, ensuring direct measurement of the absorbance properties. All measurements were conducted using unpolarized light, thereby capturing the intrinsic, isotropic absorption characteristics of the PET films. Measurements were carried out over the wavelength range of 200 to 1000 nm. The baseline was corrected using air as a reference." As the first step, before applying strain, four spectra were recorded for each sample by rotating the film to capture variations across its surface. The PET piece was rotated by 180° on each side, resulting in two measurement areas: two lines of 11 mm in length and 0.5 mm in width, located on the left and right sides of the sample center.  1 The sample was then mounted on the custom-built stretcher, subjected to a specified strain level for 30 minutes, and released. Following strain release, the same rotational approach was used to record another set of four spectra. The spectra were averaged to account for variations across the film and provide a more representative understanding of how strain affects its optical transparency. Changes in absorbance were analyzed to assess strain's impact on the films' optical properties. Figure \ref{cross-section} presents the cross-section of the PET samples before and after strain exposure. Applying strain values greater than 20\% induces distortions in the structure, making thickness measurements with an optical microscope difficult due to the appearance of shadows. The thickness of the sample was measured at different positions and an average thickness was calculated from these values (see Fig.2). It reduced from an average value of 57.271 $\mu$m to 44.342 $\mu$m responding to strain.
 
In-situ Raman spectra were obtained using the 514 nm line of an argon-ion laser (Innova 308 series, Coherent, U.S.A.) as the excitation source. A Triax 550 single monochromator (Jobin Yvon, France) equipped with 1200 grooves/mm diffraction grating was employed for spectral dispersion. The entrance slit width of the monochrometer was set to 100 µm. Broadband detection of the dispersed inelastically scattered light was achieved using liquid nitrogen-cooled CCD detectors (CCD 3500, Jobin Yvon, France). Raman spectra were recorded in a back-scattering geometry. The laser power at the sample surface was approximately 5 mW, with the laser beam focused onto the sample using an ultra-long working distance objective (50× ULWD, N.A. 0.50, Olympus, Japan), facilitating the collection of scattered light. The spectral resolution at 514 nm was 3 cm$^{-1}$. For the calibration of Raman spectrometer wave numbers, the aromatic toluene breathing mode at 1003.7 cm$^{-1}$ was used. The laser was set to have an excitation polarization parallel to the strain axis. Measurements were taken directly on the strained PET films at regular intervals of 5 minutes during the 30-minute strain application. The Raman spectra data were analyzed to extract line positions, full width at half maximum (FWHM), and intensity variations for key vibration modes. 

\section{Optical and Molecular Responses of PET Under Strain}
The interplay between mechanical strain and PET’s optical properties is a key focus of this study to gain insights into how structural changes at the molecular level translate to observable optical behavior. These implications are particularly significant for applications that balance mechanical flexibility and optical device performance, such as OSCs, flexible electronics, and strain sensors \cite{Hafsia2016}. 

The results are presented in two parts: First, UV-Vis absorption spectroscopy explores how strain affects the optical transparency of PET films, focusing on changes in absorbance across different strain levels. However, while UV-Vis reveals the transparency alterations, it does not provide insight into the underlying molecular mechanisms. To address this, in-situ Raman spectroscopy is employed to uncover the changes observed from vibration modes most sensitive to mechanical deformation, linking molecular reorganization to observed optical variations. These results are discussed in the second part.

\subsection{Strain-Induced Optical Variations}

\begin{figure}[htbp]
    \centering
    \begin{subfigure}[t]{0.5\textwidth} 
        \centering
       \includegraphics[width=\textwidth,trim=0 0 0 50,clip]{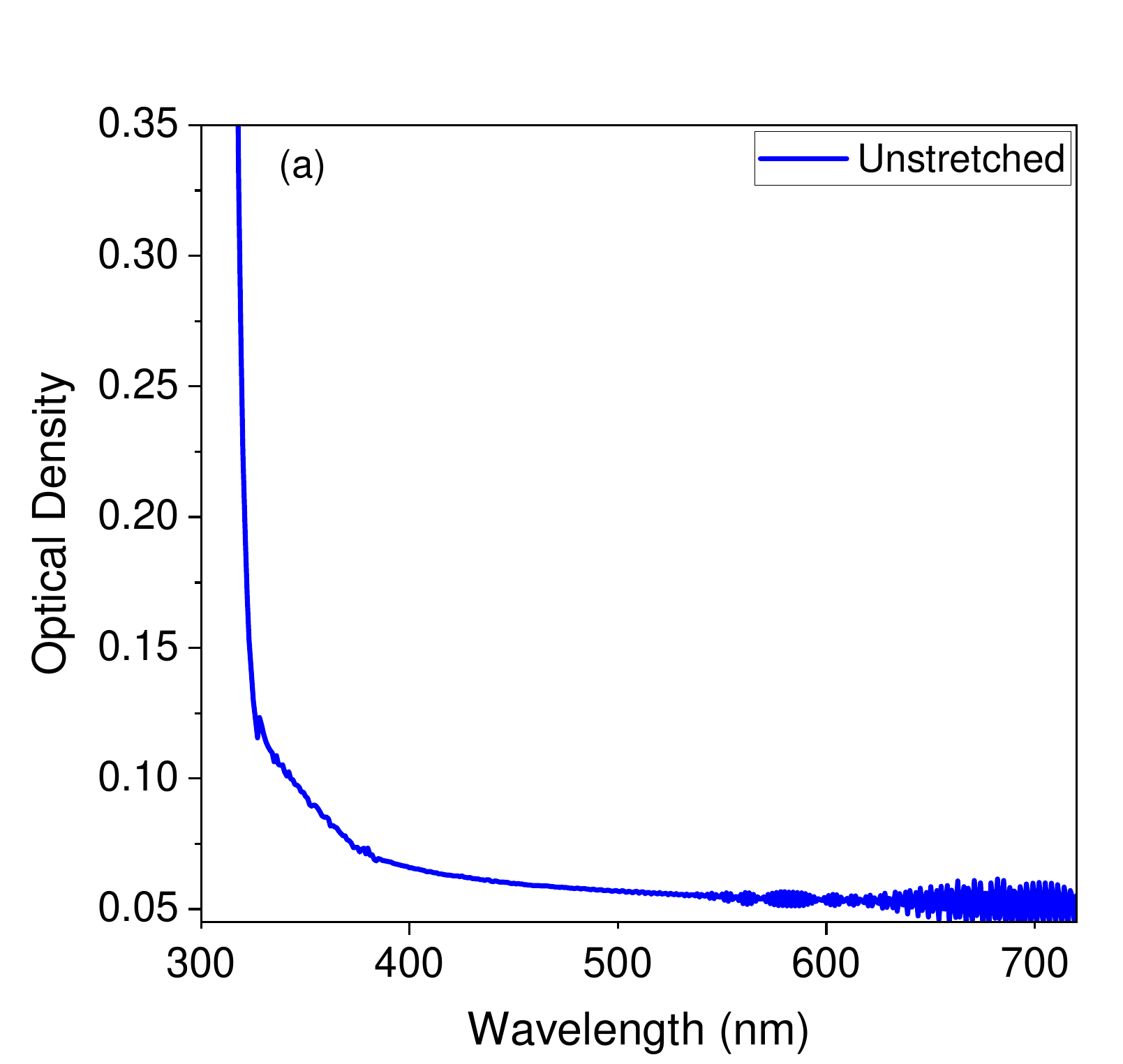}
    \end{subfigure}
   
    \begin{subfigure}[t]{0.5\textwidth}
        \centering
        \includegraphics[width=\textwidth,trim=0 0 0 40,clip]{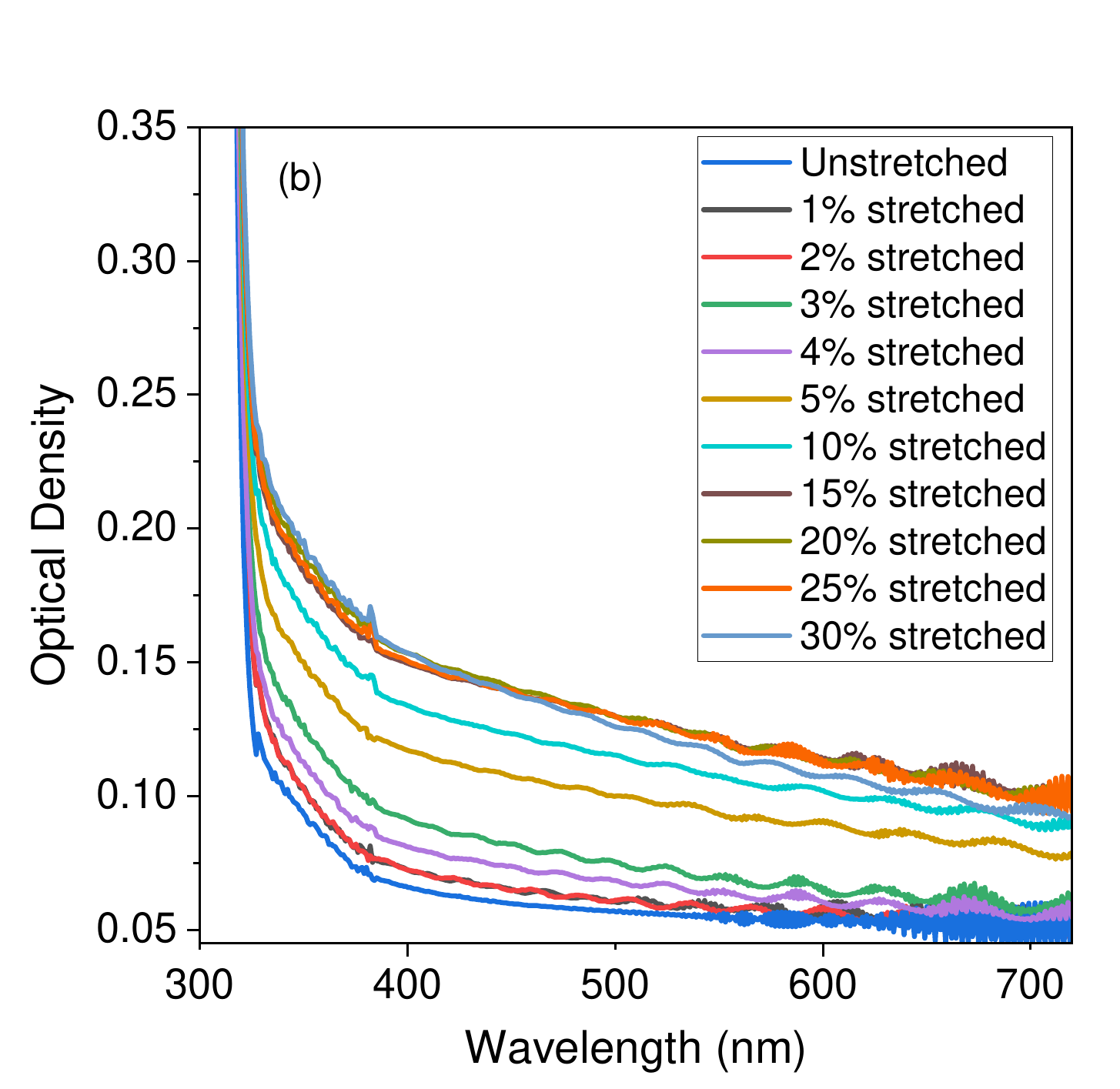}
    \end{subfigure}
    
    \caption{UV-Vis absorbance plots of PET samples: (a) unstretched PET and (b) strained PET.}
    \label{PET_all_comparison_UV-Vis}
\end{figure}

Panel (a) of Fig. \ref{PET_all_comparison_UV-Vis} shows the UV-Vis absorbance characteristics of an unstretched PET sample at room temperature. Given the transparency of PET in the visible region (400-700 nm), only minimal absorbance is observed. Panel (b) of Fig. \ref{PET_all_comparison_UV-Vis} presents the absorbance spectra of all thin-film PET samples subjected to varying strain levels after stress release. It reflects PET's typical behavior and intrinsic properties, where the absorption is stronger near the UV region and decreases for higher wavelengths, moving toward the visible spectrum. However, as the strain percentage increases from 1 to 30\%, there is a general increase in the absorbance for the UV-Vis spectrum. Increasing strain would lead to higher absorbance values for the PET films, which is particularly noticeable in the mid-range wavelengths (around 400-500 nm). This effect would interrupt the light transmission to the active layer, potentially decreasing the OSC's efficiency. 
 
\begin{figure}[h]
    \centering
        \includegraphics[width=0.5\textwidth]{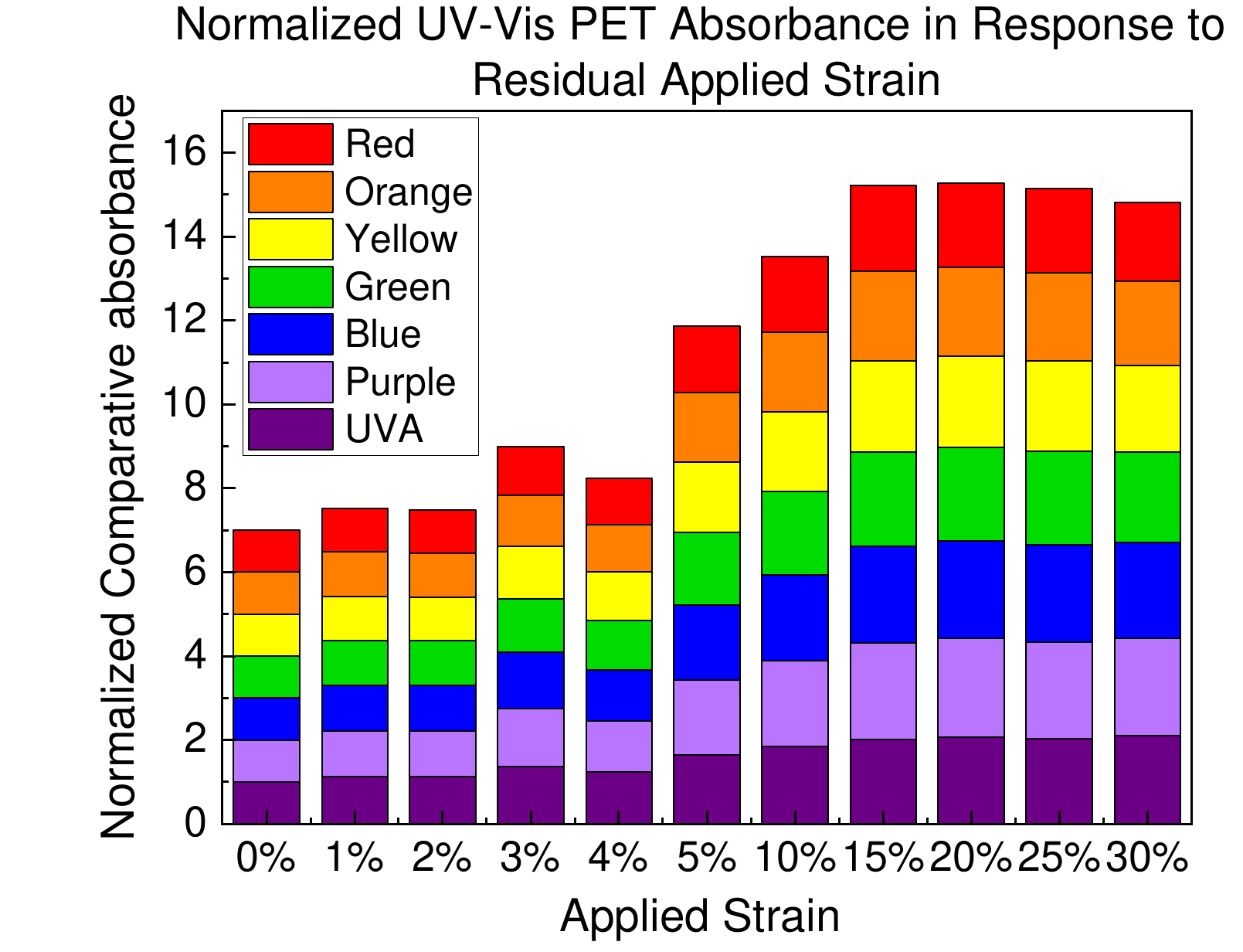}
        \caption{Normalized UV-Vis absorbance of PET films across different visible and UV spectral regions in response to varying residual applied strain. The bar graph highlights the absorbance contributions of specific wavelength ranges (UV-A, purple, blue, green, yellow, orange, and red; see also main text) at strain levels from 0 to 30\%, showing a trend of increasing absorbance with higher strain.}
        \label{bar comparison}
        \end{figure}

Figure \ref{bar comparison} illustrates the absorbance variation in the UV-vis spectrum. To produce this graph, the wavelength range between 320 and 700 nm was divided into smaller segments, each corresponding to a specific color band of the visible spectrum. The absorbance values within these ranges were summed and normalized to the unstretched values (0\%) to represent each color region's overall contribution. These regions were categorized as follows:
violet range 400–450 nm, blue range from 450–495 nm, green range from 495–570 nm, yellow range from 570–590 nm, orange range from 590–620 nm, and red range from 620–700 nm. The bars in the graph are color-coded to align with the visible spectrum, visually linking the wavelength ranges to their respective colors. This approach allowed for a clear comparison of the absorbance response of PET thin films under varying applied strains, emphasizing differences in absorbance across the spectral regions. As the strain level increases from 0\% (unstretched) to 30\%, there is a notable increase in absorbance values, especially for purple and blue wavelengths. The increase in visible light absorption can be attributed to a combination of structural changes at the molecular level. In this case, strain introduces defects or small localized energy states within the polymer, which allow PET to absorb visible light at specific wavelengths, even though it lacks an extended conjugated system. Additionally, potential variations in crystallinity and molecular alignment caused by strain create regions that scatter and refract light more effectively, reducing transparency and increasing overall absorbance.

Furthermore, there is a notable jump in absorbance values, particularly beyond 5\% strain. This indicates a possible threshold beyond which the structural changes in PET, including increased crystallinity or density of defects, lead to significantly higher light absorption (and/or scattering).

\subsection{Strain-Induced Molecular Variations Probed by Raman Spectroscopy}
In applications like OSCs where PET is used as a substrate, the observed changes in optical properties could affect the efficiency of light absorption and conversion, especially under mechanical stress. The transparency reduction in critical wavelengths is important for solar absorption since it reduces the overall performance of the solar cells. These data provide a quantitative foundation to understand how mechanical stress affects the optical properties of PET, which is crucial for both improving existing applications and developing new technologies where mechanical and optical properties intersect.
 
As mentioned above, in addition to the UV-Vis absorption measurements, we have conducted in-situ Raman spectroscopy on the stretched samples to gain deeper insights into structural alterations under applied strain. This analysis allows us to identify which molecular regions experience greater strain-induced changes and observe the PET molecule's evolution over time while exposed to external strain. We discuss the results in the following section.

\begin{figure}
\centering
\includegraphics[width=0.5\textwidth]{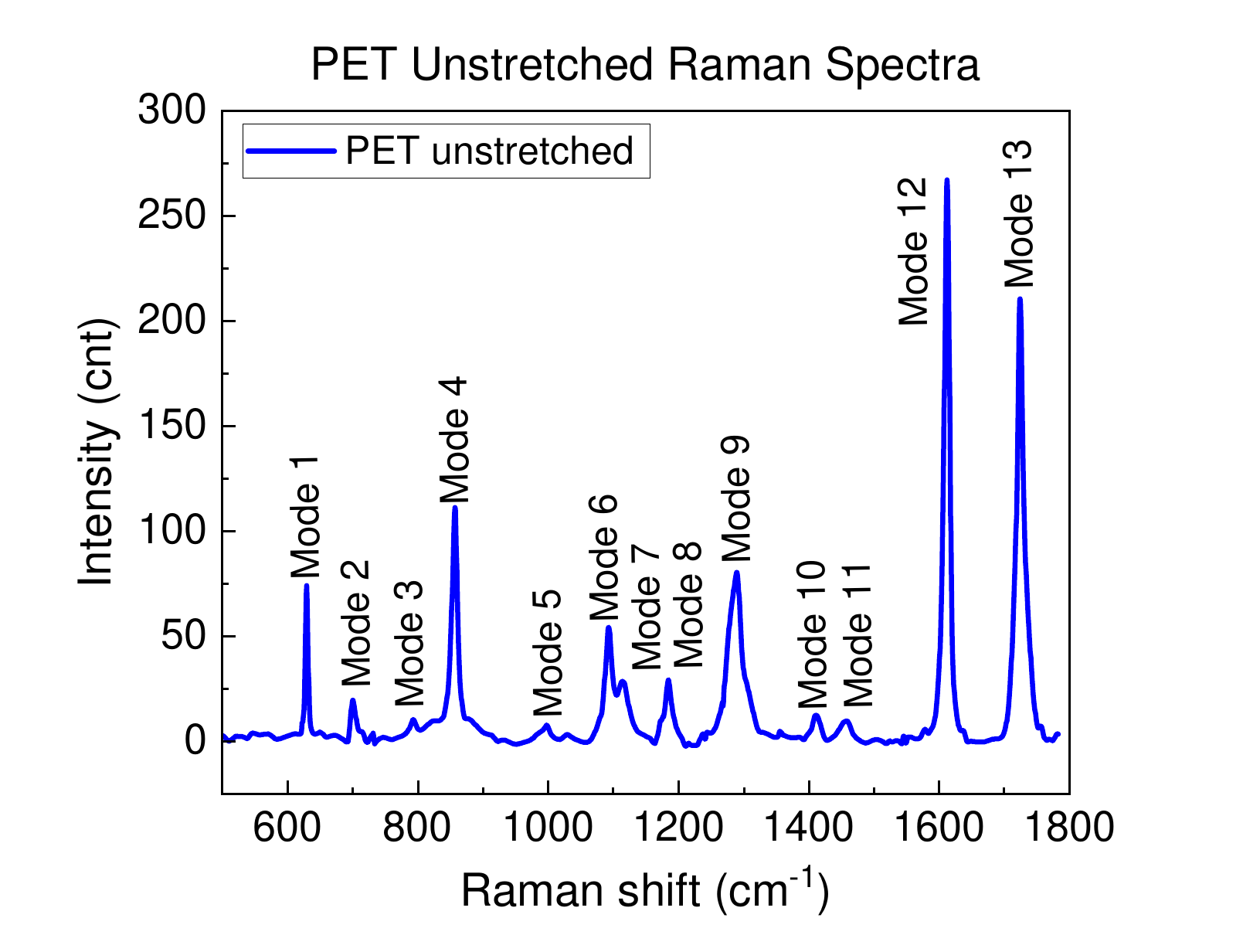}
 \caption{Raman spectrum of an unstretched PET film.} 
 \label{raman example}
\end{figure}

The Raman spectroscopy data depicts changes in the orientation of polymer chains, crystallinity, or other structural aspects. Figure \ref{raman example} shows a Raman spectrum of an unstretched piece of PET foil.

\begin{figure}[htbp]
    \centering
    \begin{subfigure}{0.3\textwidth}
        \centering 
        \caption{}
        \includegraphics[width=\textwidth,trim=0 0 5 0,clip]{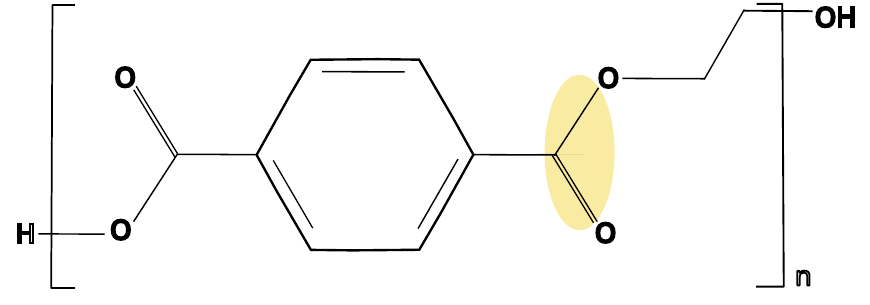}
    \end{subfigure}
    \begin{subfigure}{0.3\textwidth}
        \centering
        \caption{}
        \includegraphics[width=\textwidth,trim=0 0 5 0,clip]{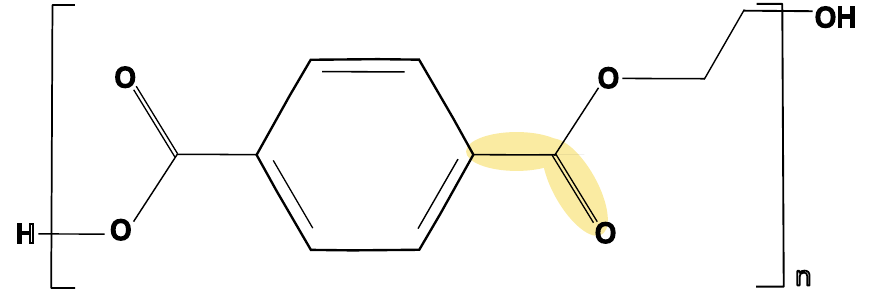}
    \end{subfigure}
    \begin{subfigure}{0.3\textwidth}
        \centering
        \caption{}
        \includegraphics[width=\textwidth,trim=0 0 5 0,clip]{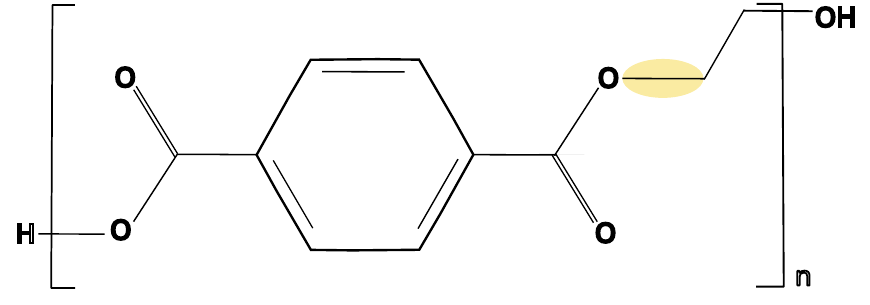}
    \end{subfigure}
    \begin{subfigure}{0.3\textwidth}
        \centering
         \caption{}
        \includegraphics[width=\textwidth,trim=0 0 5 0,clip]{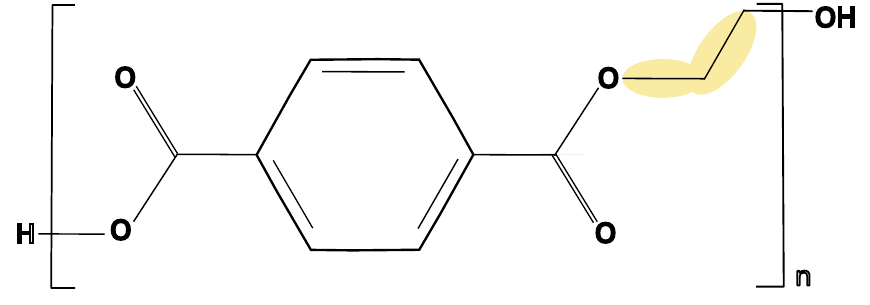}
    \end{subfigure}
    \begin{subfigure}{0.3\textwidth}
        \centering
         \caption{}
        \includegraphics[width=\textwidth,trim=0 0 5 0,clip]{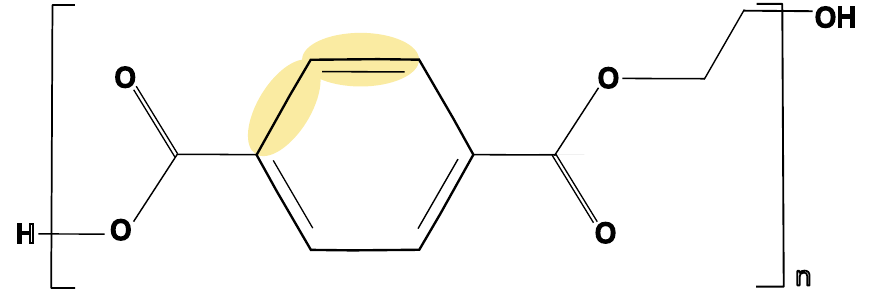}
    \end{subfigure}
    \begin{subfigure}{0.3\textwidth}
        \centering
        \caption{}
        \includegraphics[width=\textwidth,trim=0 0 5 0,clip]{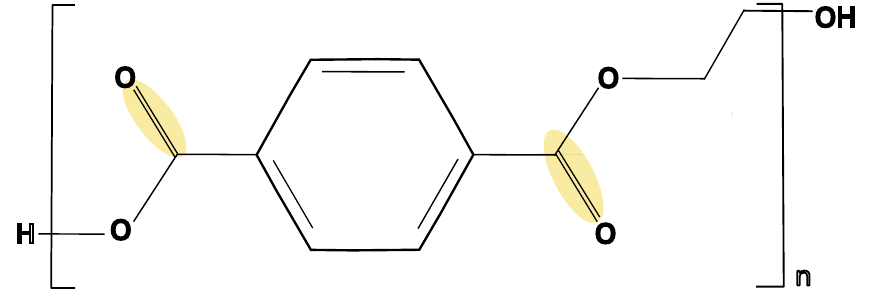}
    \end{subfigure}

    \caption{The schematics highlight the affected bonds for each mode, (a) corresponds to mode 1, (b) mode 3, (c) mode 5, (d) mode 6, (e) mode 12, and (f) mode 13.}
    \label{bond mode}
\end{figure}

\begin{table}[h!]
\scriptsize
\renewcommand{\arraystretch}{1.5} 
\centering 
\resizebox{\textwidth}{!}{%
\begin{tabular}{llll} 
\rowcolor[gray]{0.9} 
Mode order & Vibration mode                                                                 & Raman shift (1/cm) & Conformation          \\
\rowcolor[gray]{0.95} 
1          & Scissors O-C-O  \cite{cole}                                                              & $\sim$637          & -                     \\
\rowcolor[gray]{0.9} 
2          & -                                                                             & $\sim$705          & -                     \\
\rowcolor[gray]{0.95} 
3          & C=O and ring ester C-C out-of-plane bending and ring torsion \cite{Boerio, lacroix}                  & $\sim$797          & Gauche conformation \cite{cole}                     \\
\rowcolor[gray]{0.9} 
4          & Ester C(O)–O bending modes \cite{Boerio}                                               & $\sim$858          & -                     \\
\rowcolor[gray]{0.95} 
5          & C-O stretching [O-CH$_2$] \cite{lacroix}                                                 & $\sim$997          & Trans conformation    \\
\rowcolor[gray]{0.9} 
6          &  Ester C(O)–O, ethylene glycol C–C stretching \cite{rebollar, Boerio}                       & $\sim$1095         & Trans conformation    \\
\rowcolor[gray]{0.95} 
7          & Ring C–H in-plane bending, ester C(O)–O and ethylene glycol C–C stretching \cite{rebollar}   & $\sim$1119         & Gauche conformation   \\
\rowcolor[gray]{0.9} 
8          & CH$_2$ twisting mode \cite{cole}                                                         & $\sim$1186         & -                     \\
\rowcolor[gray]{0.95} 
9          &  C[O]-O stretching \cite{rebollar, Boerio}                                                      & $\sim$1292         & -                     \\
\rowcolor[gray]{0.9} 
10         & C-H bending out of the plane of the benzene ring \cite{rebollar, Boerio}                               & $\sim$1417         & -                     \\
\rowcolor[gray]{0.95} 
11         & CH$_2$ vibration bending \cite{rebollar,Boerio}                                                  & $\sim$1464         & -                     \\
\rowcolor[gray]{0.9} 
12         & C-C/C=C stretching modes \cite{Boerio}                                                    & $\sim$1613         & -                     \\
\rowcolor[gray]{0.95} 
13         & Stretching vibration C=O \cite{Boerio}                                                    & $\sim$1725         & Trans conformation                     \\
\end{tabular}%
}
\caption{Detected main modes in the Raman spectra of a PET sample}
\label{main peaks} 
\end{table}

In this spectrum, 13 main modes are identified, listed in Table \ref{main peaks}. The modes' response to applied strain is analyzed by fitting Voigt profiles to the acquired spectra (see next section). Figure \ref{bond mode} presents the demonstration of the corresponding bonds that contribute to the main Raman modes in the PET structure \cite{GonzalezCordova}.

\subsubsection{Raman Modes Wavenumber Variation in Response to Applied Strain}

Spectra were initially recorded in the unstretched state, followed by strain measurements at 5-minute intervals for 30 minutes (transitional state $T_{1}$ to $T_{6}$). Additional spectra were captured immediately after applying the strain (\( T_0 \)) as the first transition state during strain application, and after releasing the sample from stress (labeled Relea. in heatmaps), resulting in a total of 9 recorded spectra per sample, including the unstretched state (labeled Unstrt. in heatmaps). All mode shifts and FWHM values were calculated by subtracting the corresponding unstretched values to assess the deviations. For intensity measurements, the data were normalized to the unstretched state for each sample to better understand the variations induced by stretching.

The Raman modes of the molecules in the sample correspond to specific atomic movements within the molecular structure, such as stretching, bending, twisting, or rocking of molecular bonds (see Table 1). The line positions indicate the energy of the vibration mode, which gives information on the strength of the bonds and the changed interactions \cite{Jones2019}.

\begin{figure}[htbp]
    \centering
    \begin{subfigure}{0.43\textwidth}
        \centering 
        \caption{}
        \includegraphics[width=\textwidth, trim=0 0 0 0,clip]{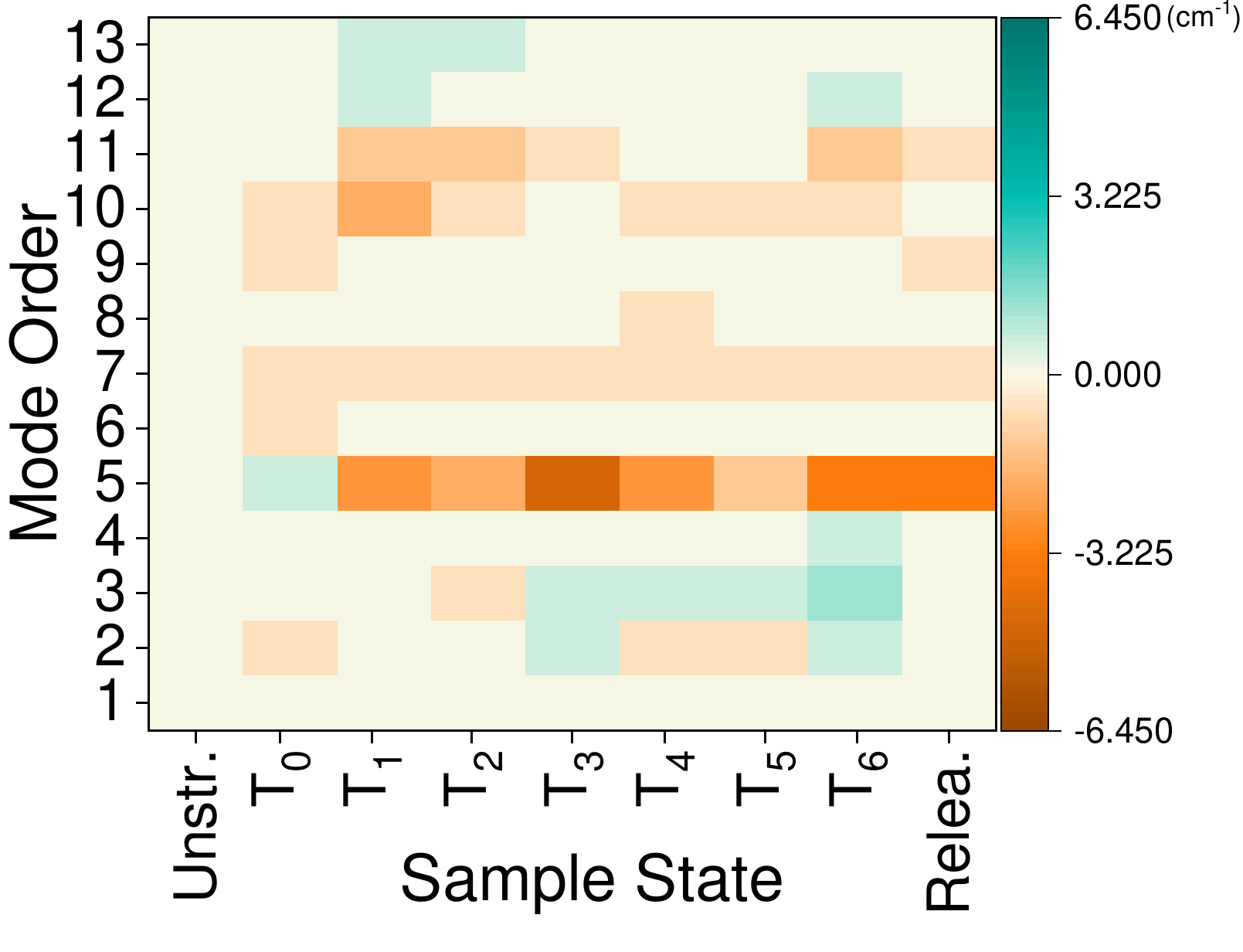}
    \end{subfigure}
    \hspace{0.02\textwidth} 
    \begin{subfigure}{0.43\textwidth}
        \centering
        \caption{}
        \includegraphics[width=\textwidth,trim=0 0 50 10,clip]{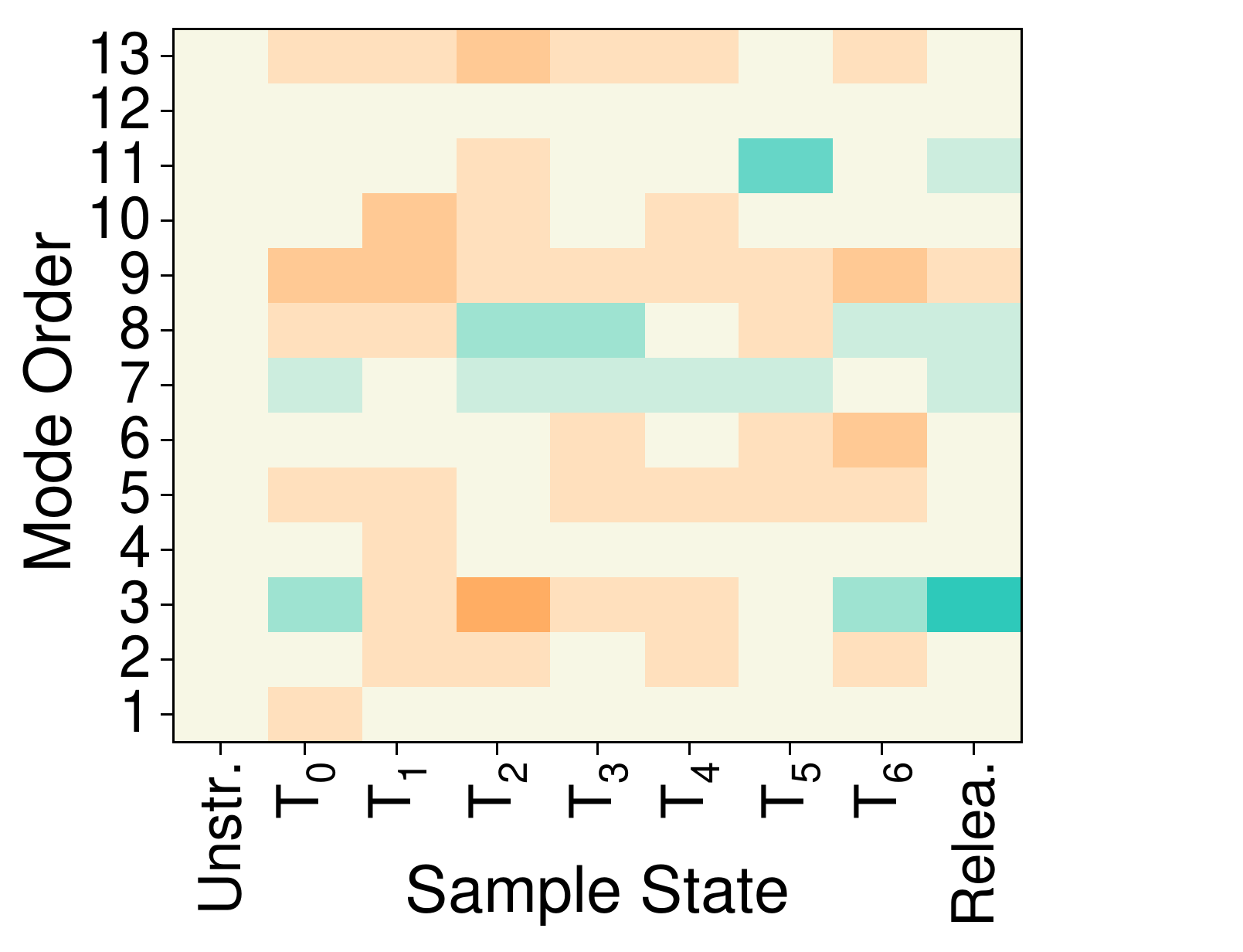}
    \end{subfigure}
    \vspace{-0.1cm} 

    \begin{subfigure}{0.43\textwidth}
        \centering
         \caption{}
        \includegraphics[width=\textwidth,trim=15 0 60 0,clip]{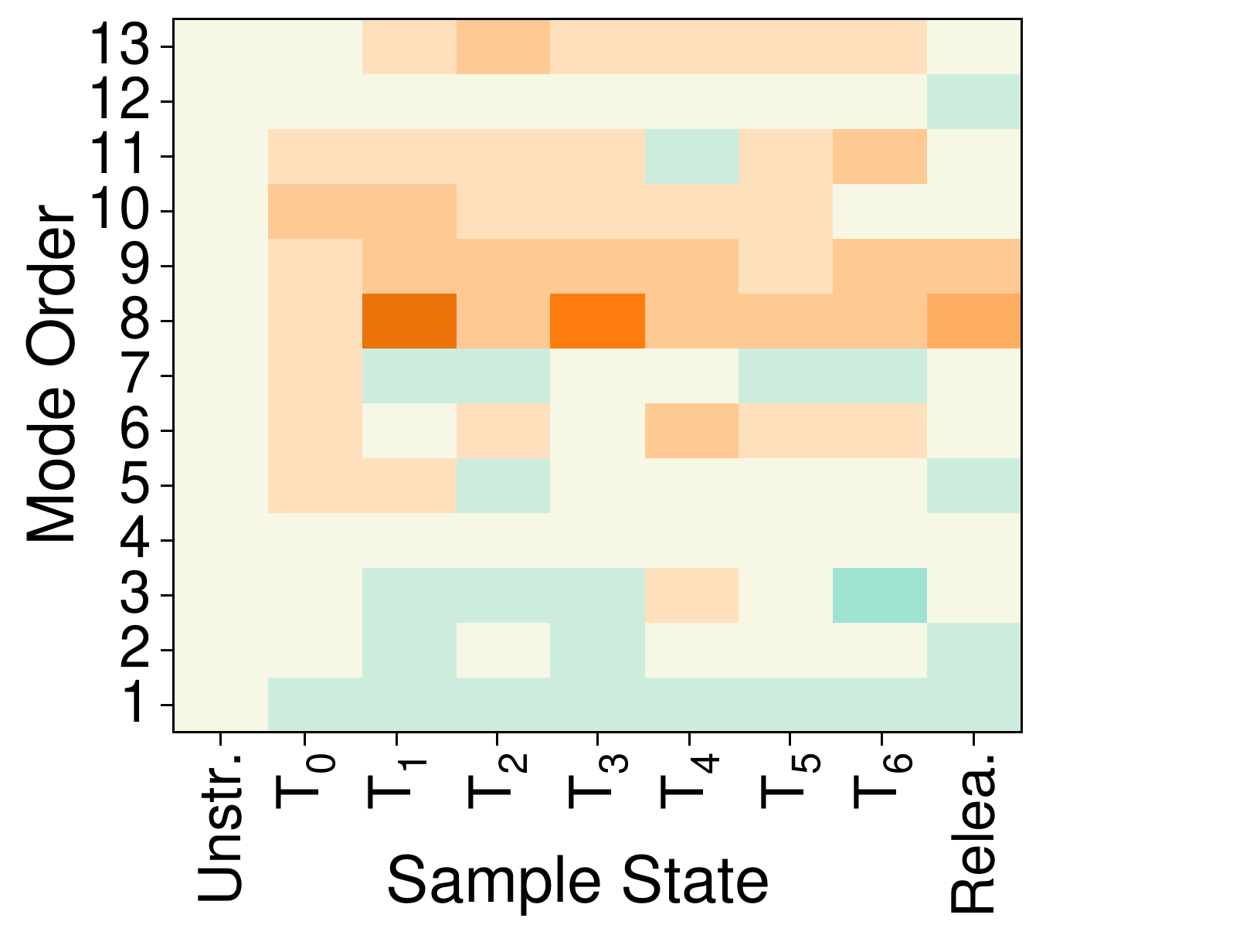}
    \end{subfigure}
    \hspace{0.02\textwidth} 
    \begin{subfigure}{0.43\textwidth}
        \centering
        \caption{}
        \includegraphics[width=\textwidth,trim=0 0 70 0,clip]{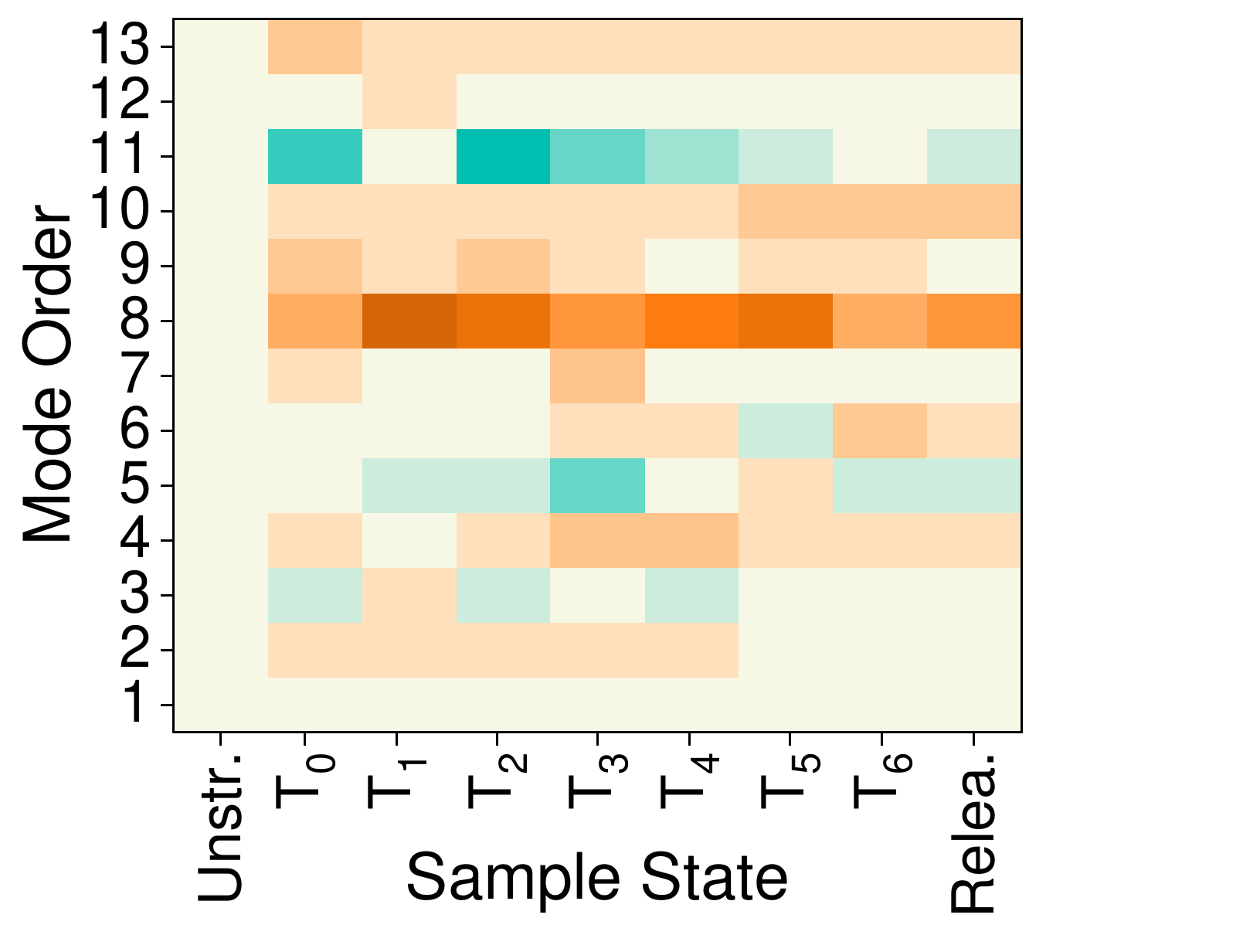}
    \end{subfigure}
    \vspace{-0.1cm} 

    \begin{subfigure}{0.43\textwidth}
        \centering
         \caption{}
        \includegraphics[width=\textwidth,trim=15 0 60 0,clip]{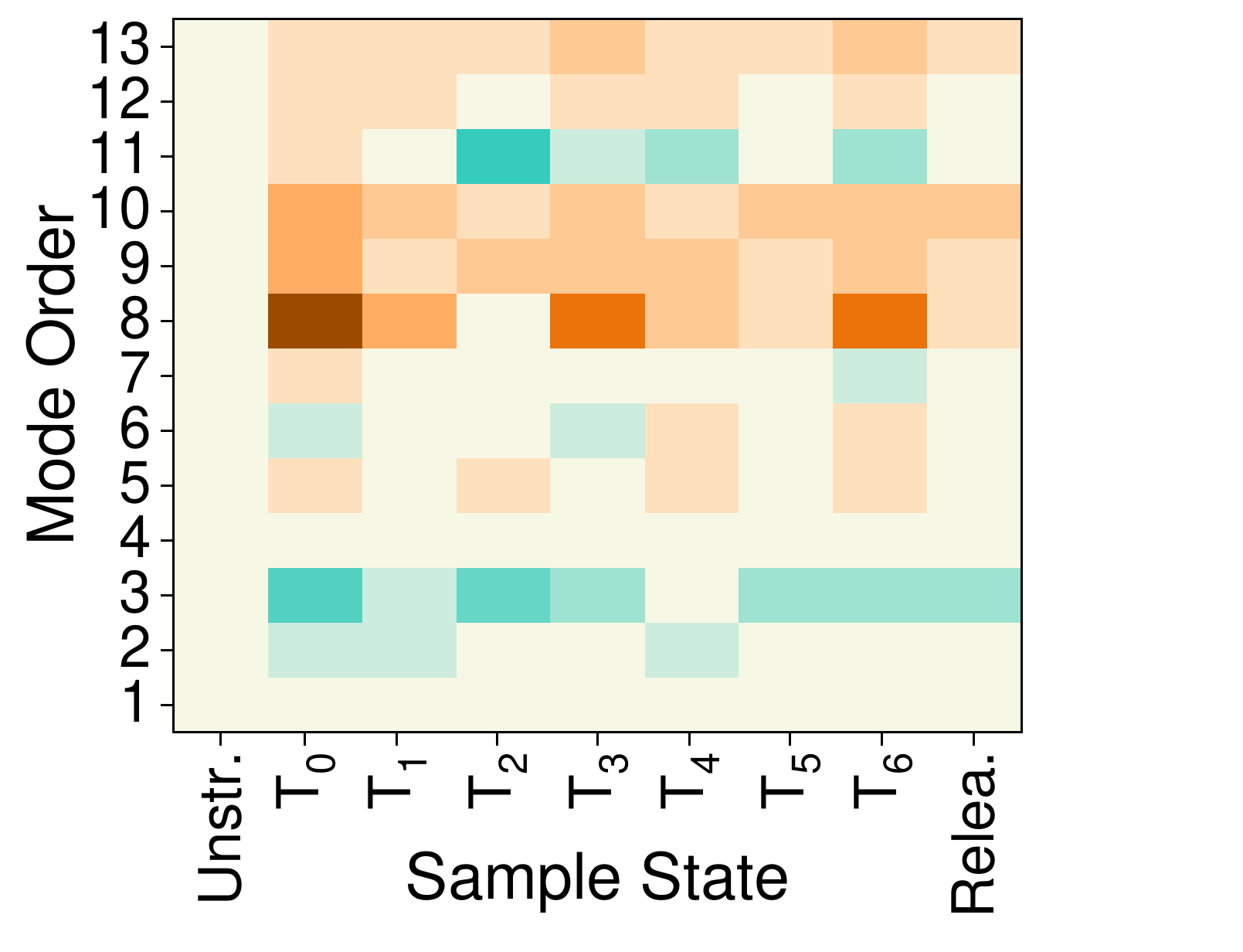}
    \end{subfigure}
    \hspace{0.02\textwidth} 
    \begin{subfigure}{0.43\textwidth}
        \centering
        \caption{}
        \includegraphics[width=\textwidth,trim=0 0 60 0,clip]{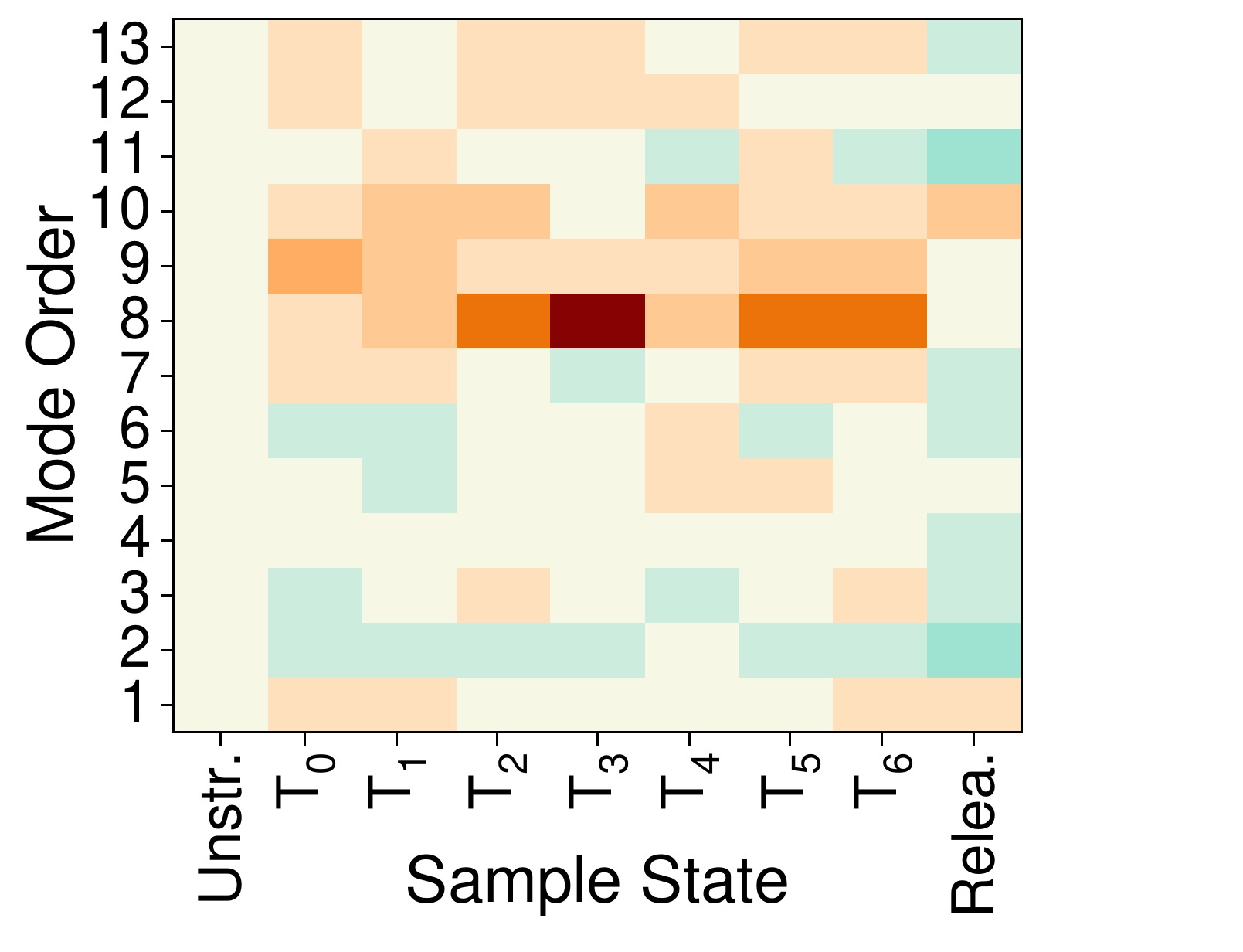}
    \end{subfigure}

    \caption{The graphs present the state evolution of relative Raman spectra mode positions for PET under (a) 1\% (b) 5\%, (c) 10\%, (d) 15\%, (e) 20\%, and (f) 25\% applied tensile strain. The heatmaps compare the mode positions at various transitional states (x-axis) from $T_{0}$ to $T_{6}$ during strain application and after strain release (Relea.) The color bar is unified across different heatmaps. For each sample state, the measured value of each mode is subtracted from the corresponding value in the unstretched state.}
    \label{mode_all_heatmap}
\end{figure}

Figure \ref{mode_all_heatmap} illustrates the variation of each line in response to strain across different transitional states. A uniform scale was applied across all heatmaps to ensure comparability. The results reveal that the behavior of the lines varies significantly, some exhibit only minor changes, while others undergo substantial variations when subjected to strain. 

At low strain levels (1 to 4\%), deviations in Raman mode positions are minimal and localized, affecting mainly specific modes such as 5 and 7 (see Table \ref{main peaks}), corresponding to C-O stretching [\( \mathrm{O}-\mathrm{CH}_2 \)] and C-C stretching of the ethylene glycol unit. These vibration modes show high susceptibility to mechanical deformation. However, after releasing the strain, most modes return to their original positions, indicating elastic behavior with high reversibility. As the strain increases from 1\% to 4\%, slightly larger shifts are observed, but the changes remain constrained to a few modes. No significant structural changes are detected, and the material retains its molecular integrity. This reversible behavior reflects the resilience and ability of the material to absorb minor mechanical stress without significant structural disruption. 

For the moderate strain range (5 to 10\%), the Raman wavenumber shifts become more pronounced, as residual shifts persist after strain release. These shifts suggest the onset of molecular reconfiguration, which gradually transitions to more noticeable structural effects. By 5\%, the deviations have stabilized across more modes, with cumulative structural effects becoming evident. The eighth mode, associated with CH$_2$, shows particular sensitivity, marking the transition toward localized but more significant molecular changes. At 10\% strain, the shifts become more widespread, impacting multiple modes such as C=O and ring ester C-C out-of-plane bending and ring torsion (mode 3), ethylene glycol C–C and C-O stretching (mode 6), CH$_2$ twisting (mode 8), and the ester O-C stretching (mode 9). The deviation range narrows slightly in comparison to smaller strain values but becomes more uniformly distributed across most modes, indicating that structural changes are no longer localized. The return to baseline positions after strain release is increasingly limited. This molecular reorganization is accompanied by partial anisotropy, altering optical properties such as transparency and birefringence (see UV-Vis response). Localized molecular alignment as well as the creation of defects would lead to light scattering and reduced transparency. Modes such as the ninth and eleventh (O-C stretching), which have shown minimal deviations at lower strains, now exhibit significant shifts, suggesting broader structural disruption.

At higher strain levels (15\% to 30\%), the shifts become extreme and persistent, with residual deviations observed across nearly all modes. Modes 5 and 8 consistently show the strongest responses, reflecting their high sensitivity to strain. Other modes, such as the third, ninth, and eleventh, also exhibit significant shifts. Higher color contrast in the corresponding heatmaps highlights the extensive and permanent structural changes, with almost no reversibility after strain release.  At 30\% strain, the material is exhibiting a complete transition from elastic to plastic behavior. Table \ref{comprehensive peaks} presents a comprehensive overview of the variation in the position of the modes in response to applied stress. \begin{table}[H]
\scriptsize
\centering
\begin{tabularx}{\textwidth}{|l|X|X|X|X|}
\hline
\rowcolor[gray]{0.8} 
\textbf{Strain Level} & \textbf{Key modes Affected} & \textbf{Shift Patterns} & \textbf{Reversibility After Release} & \textbf{Conclusion} \\ \hline
\rowcolor[gray]{0.9} 
1\% & 5, 7 & Large shifts, localized & High reversibility & elastic behavior with minimal residual effects. \\ \hline
\rowcolor[gray]{0.95} 
2\% & 5, 7, 10 & minor shifts, more modes affected & Partial reversibility & Increased sensitivity of modes; minimal residual effect. \\ \hline
\rowcolor[gray]{0.9} 
3\% & 5, 7, 11 & Moderate shifts, more spread across modes & Partial return to baseline & Further deviations, signs of increased sensitivity. \\ \hline
\rowcolor[gray]{0.95} 
4\% & 5, 7, 10, 11 & Noticeable shift increase with broader deviations & Limited return & Higher structural sensitivity with noticeable changes in mode behavior. \\ \hline
\rowcolor[gray]{0.9} 
5\% & 3, 5, 8 & minor shifts, spread across most modes & Residual shifts remain & Signs of permanent changes, increased structural alterations. \\ \hline
\rowcolor[gray]{0.95} 
10\% & 3, 6, 8, 9 & Consistent shifts across multiple modes & Strong residual effects & Non-reversible response, extensive molecular changes start. \\ \hline
\rowcolor[gray]{0.9} 
15\% & 8, 10, 11 & Broader, strong deviations across many modes & Very limited return & High level of alteration with dominant irreversible effects. \\ \hline
\rowcolor[gray]{0.95} 
20\% & 3, 8, 9, 10, 11 & Large deviations, intense mode shifts & No recovery observed & Severe, likely permanent deformation across molecular structure. \\ \hline
\rowcolor[gray]{0.9} 
25\% & 8, 9, 10 & Consistent deviations across all modes & No recovery & Extensive reorganization, clear irreversible changes. \\ \hline
\rowcolor[gray]{0.95} 
30\% & All modes & Broad deviations, complete departure & No recovery & Permanent structural alteration, total loss of elasticity. \\ \hline
\end{tabularx}
\caption{Strain-level effects on Raman line wavenumber positions}
\label{comprehensive peaks}
\end{table}

\subsubsection{Raman modes FWHM Variation in Response to Applied Strain}
The shapes of Raman lines are typically described by Voigt profiles, which are convolutions of Gaussian and Lorentz line shapes. This also considers inhomogeneous broadening mechanisms. In the case of a solid sample, the degree of FWHM broadening is often associated with structural disorder or heterogeneity in the material. As already reflected in the observed wavenumber shifts of the different vibrational modes, strain can introduce various local distortions, such as changes in the lengths or angles of the bonds within the polymer structure \cite{Jones2019}. These local variations could shift the mode and broaden it, as the material under strain exhibits a range of slightly different vibration wavenumber due to the nonuniform stress distribution.

Since strain often affects each molecular environment differently, the FWHM is sensitive to these subtle differences which can result in larger variations compared to wavenumber mode shifts. FWHM captures the “spread” of stress-induced vibration states while wavenumber shifts reflect overall bond stiffening or softening and slight bond length adjustment, which leads to a shift in the vibration wavenumber. However, this effect is generally more uniform and subtle, reflecting an average change in bond lengths across the material rather than the localized variations contributing to broadening. Therefore, while the wavenumber shift shows a consistent structural adjustment under strain, it does not capture the localized disorder as effectively as FWHM, which is more sensitive to minor variations in molecular structure in response to strain \cite{Angel} These structural changes are of importance since they can strongly affect the electronic properties of the samples as has been demonstrated earlier \cite{Materny2001425}.

\begin{figure}[htbp]
    \centering
    \begin{subfigure}{0.43\textwidth}
        \centering 
        \caption{}
        \includegraphics[width=\textwidth, trim=0 0 0 0,clip]{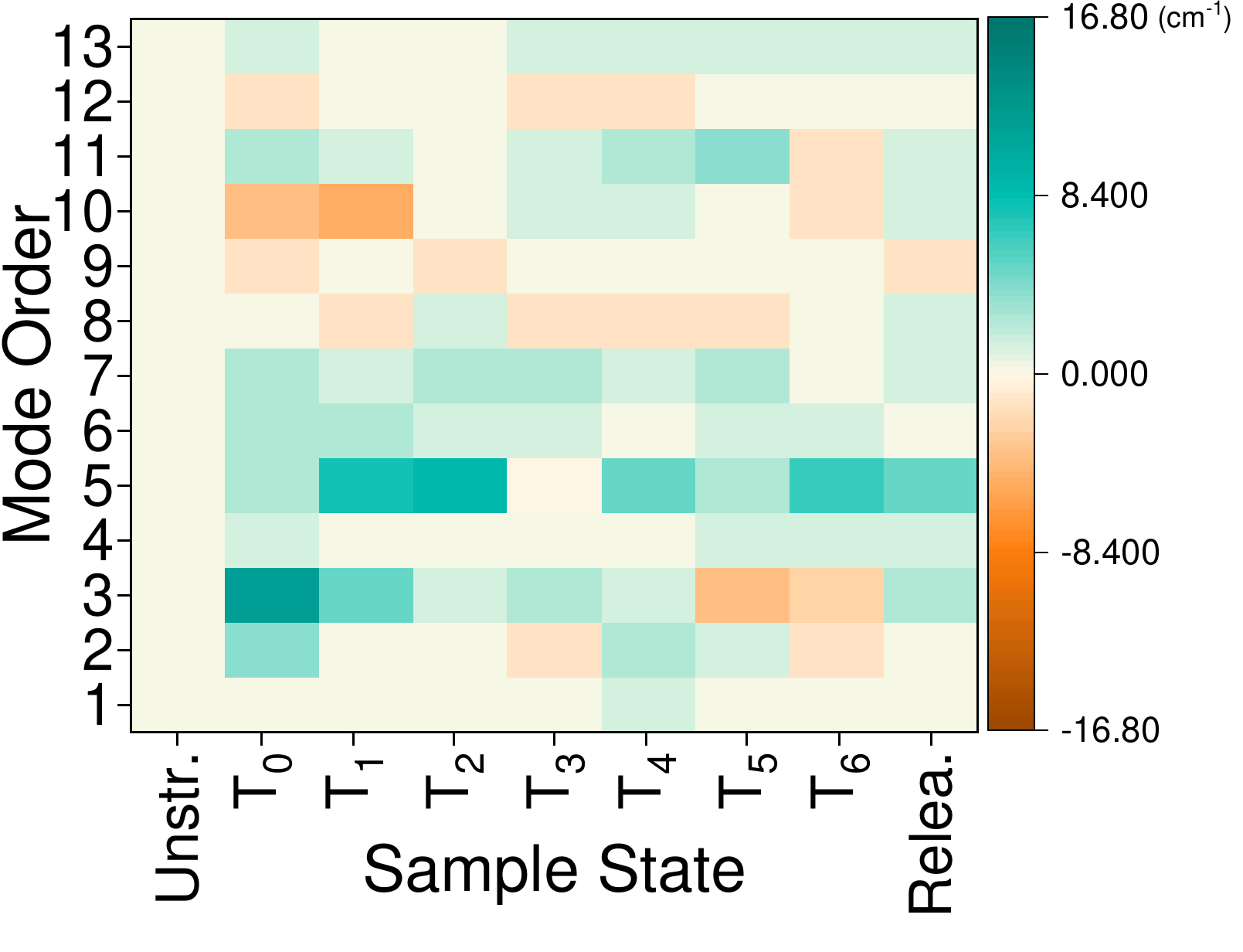}
    \end{subfigure}
    \hspace{0.02\textwidth} 
    \begin{subfigure}{0.43\textwidth}
        \centering
        \caption{}
        \includegraphics[width=\textwidth,trim=-5 0 50 10,clip]{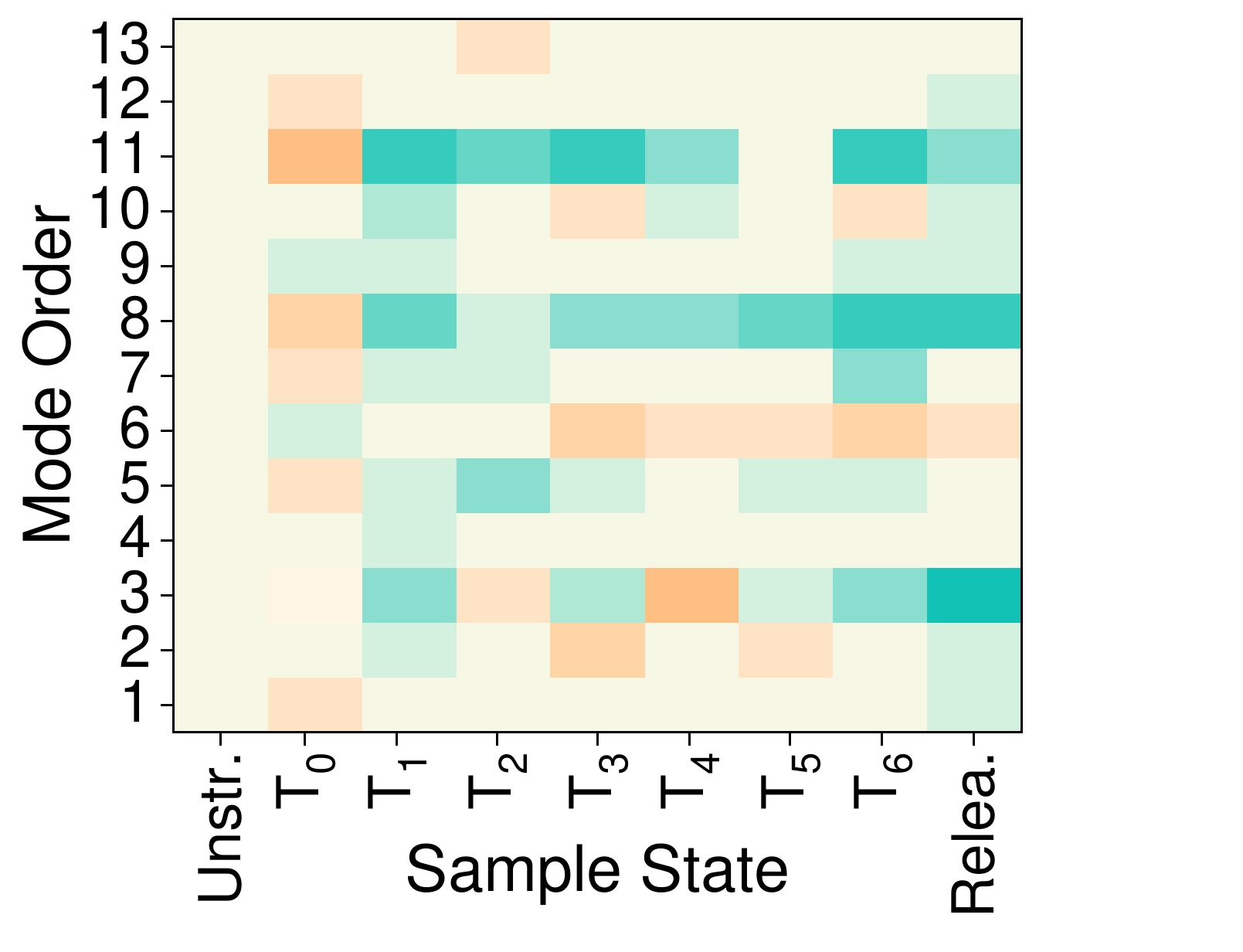}
    \end{subfigure}
    \vspace{-0.1cm} 

    \begin{subfigure}{0.43\textwidth}
        \centering
         \caption{}
        \includegraphics[width=\textwidth,trim=10 0 60 0,clip]{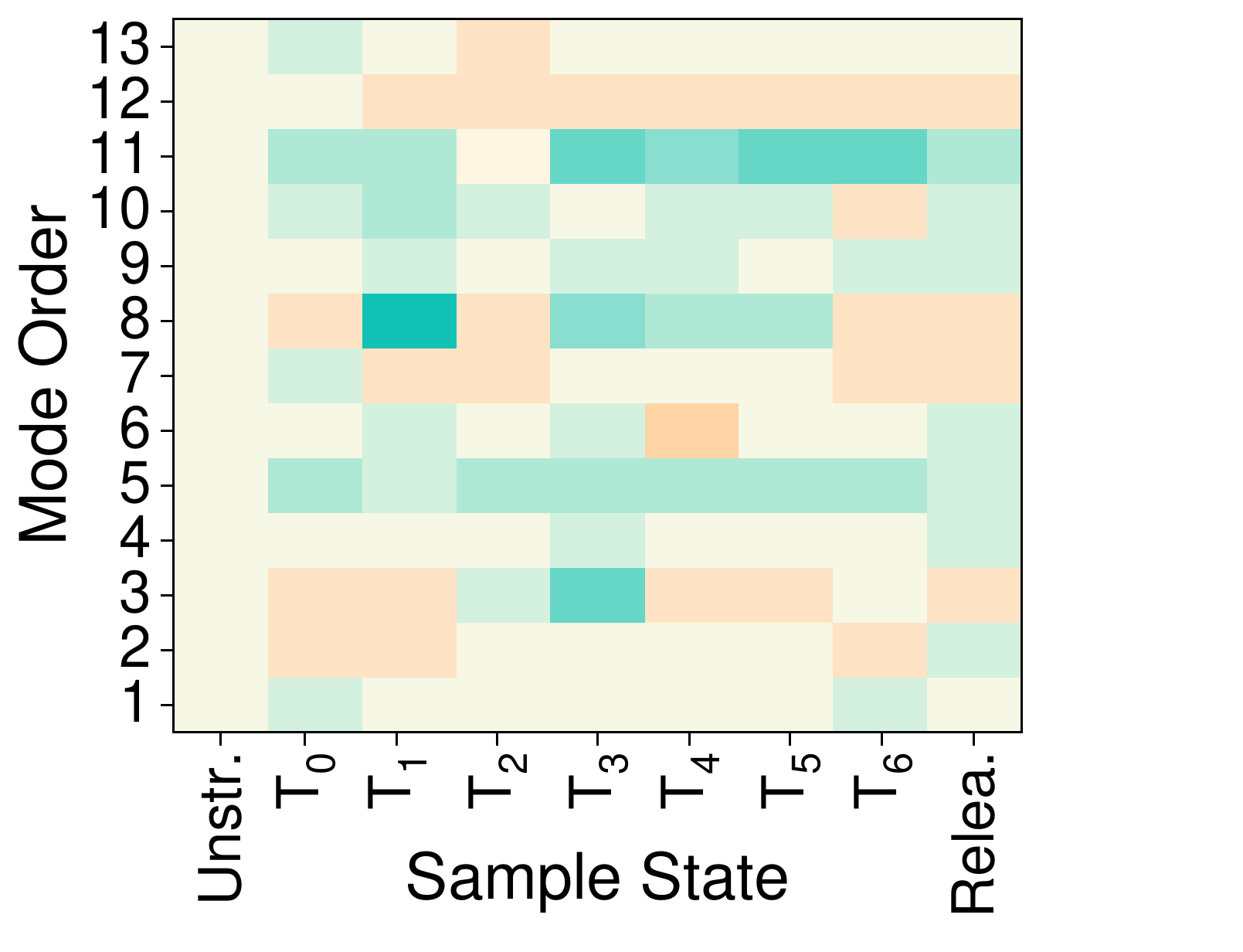}
    \end{subfigure}
    \hspace{0.02\textwidth} 
    \begin{subfigure}{0.43\textwidth}
        \centering
        \caption{}
        \includegraphics[width=\textwidth,trim=0 0 70 0,clip]{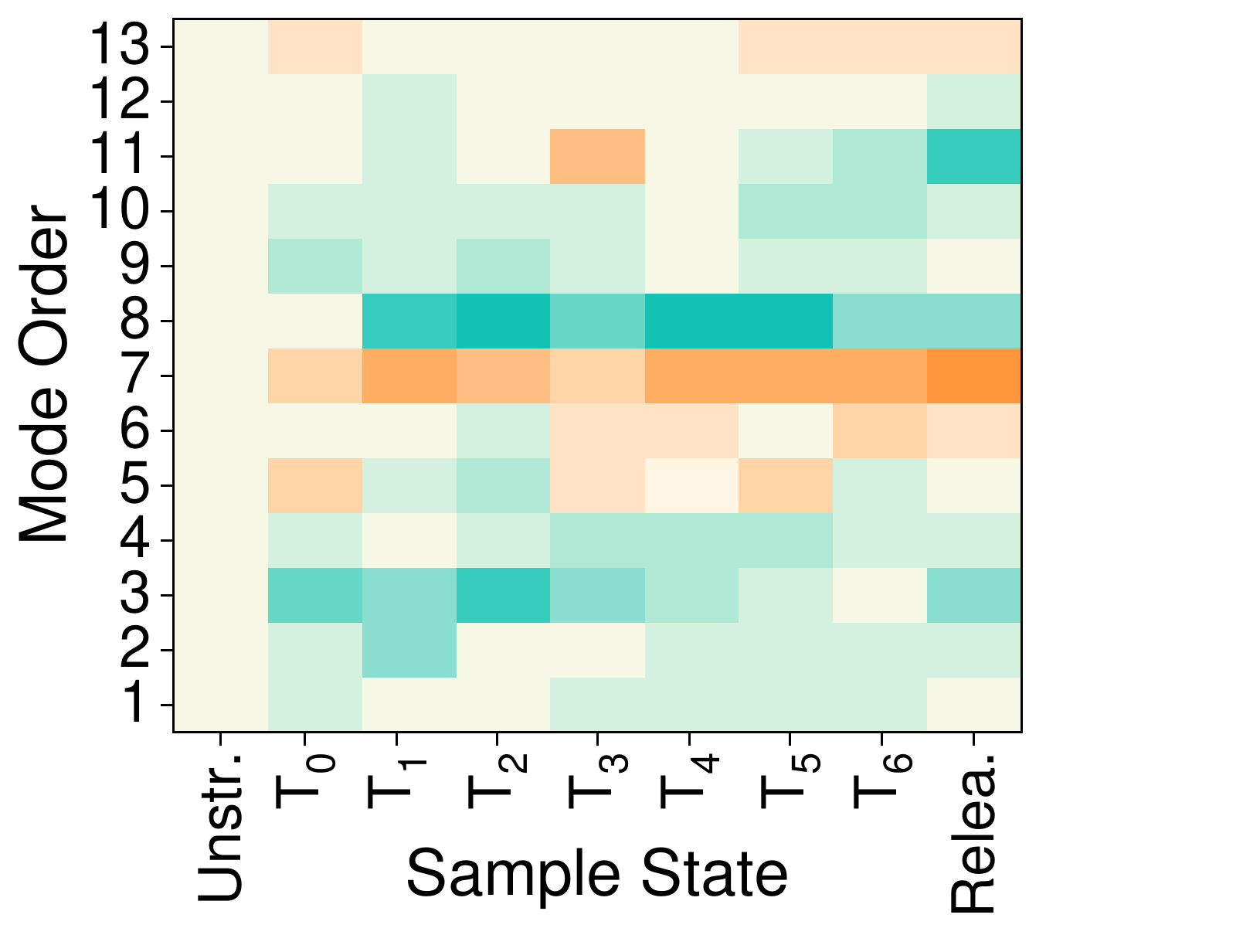}
    \end{subfigure}
    \vspace{-0.1cm} 

    \begin{subfigure}{0.43\textwidth}
        \centering
         \caption{}
        \includegraphics[width=\textwidth,trim=10 0 60 0,clip]{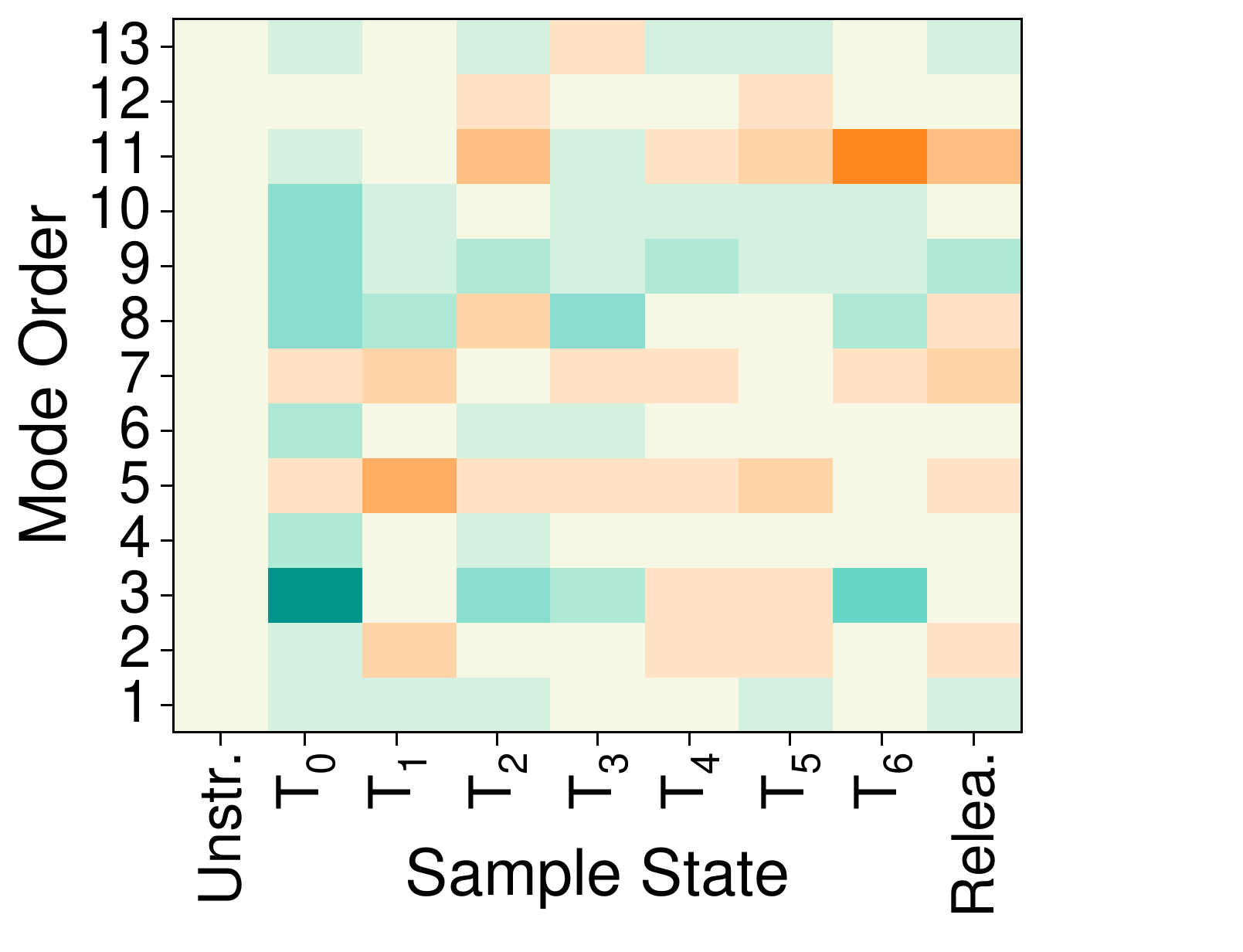}
    \end{subfigure}
    \hspace{0.02\textwidth} 
    \begin{subfigure}{0.43\textwidth}
        \centering
        \caption{}
        \includegraphics[width=\textwidth,trim=0 0 60 0,clip]{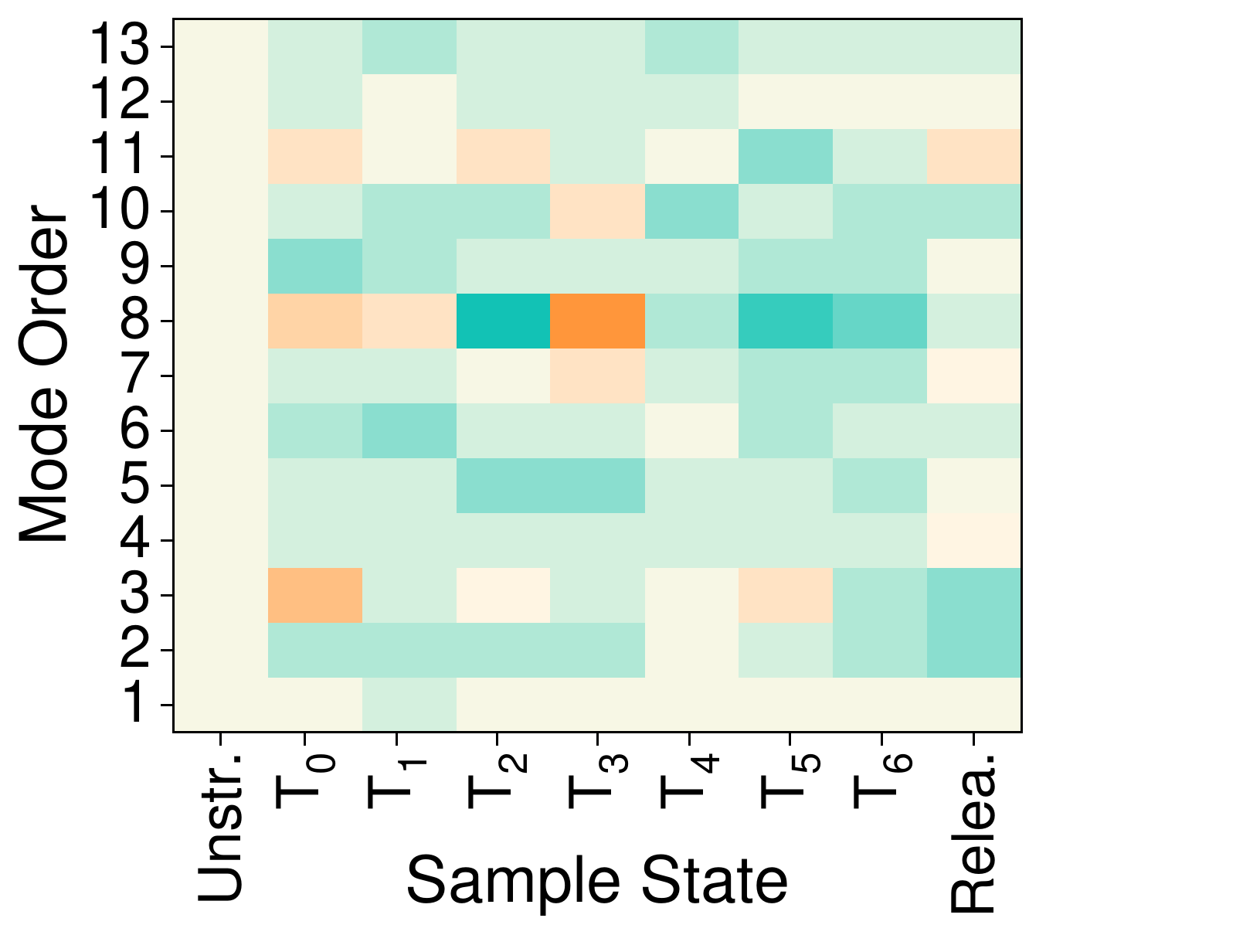}
    \end{subfigure}

    \caption{The graphs present the state evolution of relative Raman mode's FWHM variations for PET under (a) 1\% (b) 5\%, (c) 10\%, (d) 15\%, (e) 20\%, and (f) 25\% applied tensile strain. The heatmaps compare the mode's FWHM at various transional states (x-axis) from $T_{0}$ to $T_{6}$ during strain application and after strain release (Relea.) The color bar is unified across different heatmaps. For each sample state, the measured value of each mode is subtracted from the corresponding value in the unstretched state.}
    \label{chsoen strain-time FWHM maps}
\end{figure}

Figure \ref{chsoen strain-time FWHM maps} shows the variation of FWHM response to different strain values. For a small amount of strain (1 to 4\%) relatively strong broadening of FWHM is observed, which is primarily localized to the modes 3 and 5.
Broadening stabilizes early around T\(_{4}\) and show minimal recovery after release. 

For midrange strain values (5-10\%) a noticeable broadening begins across most modes. At the 5\% stretch, it has intensified across modes 3, 8, and 11. Persistent broadening after T\(_{4}\)  with partial recovery occurs after release. This stage can be considered as the initial step for the material to begin its transition from elastic to plastic deformation. Stress spreads across various modes, and strain-induced heterogeneity develops. For 10\% stretch, FWHM values remain elevated with minimal but consistent residual effects. showing further structural sensitivity and permanent changes.

For higher strain levels (15 to 30\%), broadening affects all modes, with little to no recovery after release. At 30\%, broadening is widespread and consistent. Here, irreversible deformation dominates, with molecular dislocations and reorganizations becoming widespread. Loss of elasticity and permanent structural changes are observable. Table \ref{table FWHM all} presents a comprehensive overview of the response to the FWHM variations in the observed mode.\begin{table}[h!]
\scriptsize
\centering
\begin{tabularx}{\textwidth}{|l|X|X|X|X|}
\hline
\rowcolor[gray]{0.8} 
\textbf{Strain Level} & \textbf{Key modes Affected} & \textbf{Shift Patterns} & \textbf{Reversibility After Release} & \textbf{Conclusion} \\ \hline
\rowcolor[gray]{0.9} 
1\% & 3 5 & large shifts, localized & High reversibility & Elastic behavior with minimal residual effects. \\ \hline
\rowcolor[gray]{0.95} 
2\% & 3, 5, 7 & moderate shifts and mainly localized & Partial reversibility & Early signs of structural sensitivity to strain. \\ \hline
\rowcolor[gray]{0.9} 
3\% & 3, 5, 11 & moderate shifts and mainly localized & Partial return to baseline & Increased molecular perturbations, sensitivity rising. \\ \hline
\rowcolor[gray]{0.95} 
4\% & 3, 5, 7, 11 & noticeable shifts & Limited return & Enhanced structural changes, limited elasticity. \\ \hline
\rowcolor[gray]{0.9} 
5\% & 3, 8, 11 & moderate shifts & Residual broadening remains & Onset of permanent structural alterations. \\ \hline
\rowcolor[gray]{0.95} 
10\% & 3, 5, 8, 11 & Broad, minimal deviations & Strong residual effects & Major irreversible alterations with dominant effects. \\ \hline
\rowcolor[gray]{0.9} 
15\% & 3, 5, 7, 8 & Broad, with strong deviations for certain peaks & Very limited return & Major irreversible alterations with dominant effects. \\ \hline
\rowcolor[gray]{0.95} 
20\% & Nearly all modes affected & small shifts across all modes & No recovery observed & Permanent deformation, structural reorganization. \\ \hline
\rowcolor[gray]{0.9} 
25\% & Nearly all modes affected & Severe, consistent deviations & No recovery & Extensive reorganization, clear irreversible changes. \\ \hline
\rowcolor[gray]{0.95} 
30\% & All modes & moderate shifts and broadening across all modes & No recovery & Permanent structural alteration, total loss of elasticity. \\ \hline
\end{tabularx}
\caption{Summary of strain-level effects and their impact on Raman mode FWHM broadening}
\label{table FWHM all}
\end{table}

\subsubsection{Raman Modes Intensity Variation in Response to Applied Strain}
The Raman mode intensity is related to the polarizability tensor, which describes how the electron cloud in a molecule deforms in response to incident light. Strain application would cause a change in the Raman intensity due to an alteration in the molecular symmetry of the molecule or crystalline parts of the polymer structure, which affects specific elements of this tensor. In our study, the laser excitation polarization is fixed and parallel to the strain axis, therefore, only a subset of the tensor elements is being probed. Typically, with a fixed polarization parallel to the strain axis, the diagonal components of the polarizability tensor (e.g., $\alpha_{xx}$, $\alpha_{yy}$, $\alpha_{zz}$), (depending on how the axes are defined) dominate the response. Off-diagonal terms (e.g., $\alpha_{xy}$) require changes in polarization or more complex experimental setups to be observed \cite{Jones2019}.

By focusing on a fixed polarization, this study has isolated and examined the strain effects of the Raman-active modes directly along the strain axis. This provides a clear and reproducible dataset without adding the complexity of polarization rotation. Since we have already discussed wavenumber shifts and FWHM variations, adding intensity data completes the picture of the strain impacts on the Raman spectrum.
\begin{figure}[htbp]
    \centering
    \begin{subfigure}{0.43\textwidth}
        \centering 
        \caption{}
        \includegraphics[width=\textwidth, trim=10 0 5 0,clip]{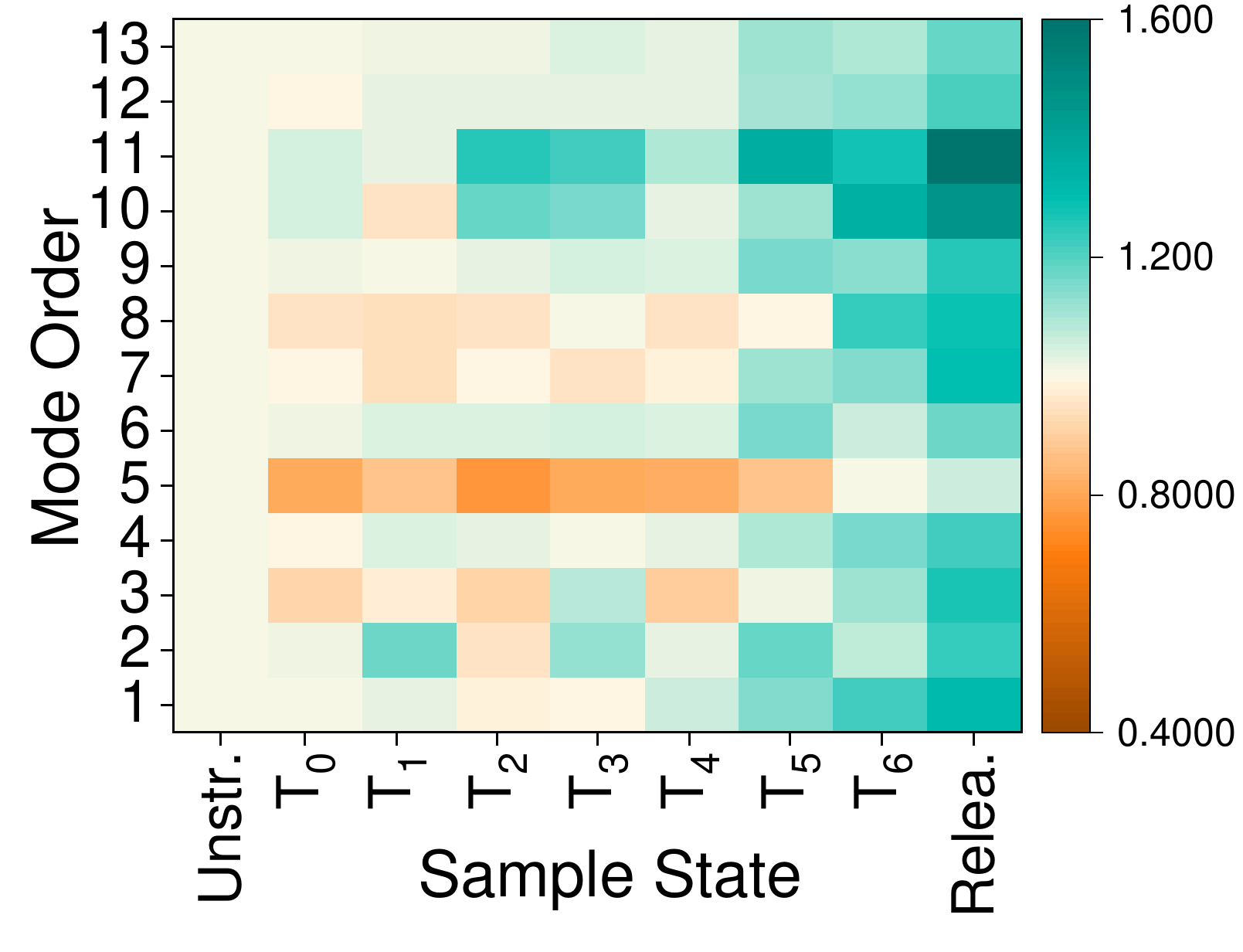}
    \end{subfigure}
    \hspace{0.02\textwidth} 
    \begin{subfigure}{0.43\textwidth}
        \centering
        \caption{}
        \includegraphics[width=\textwidth,trim=0 0 50 10,clip]{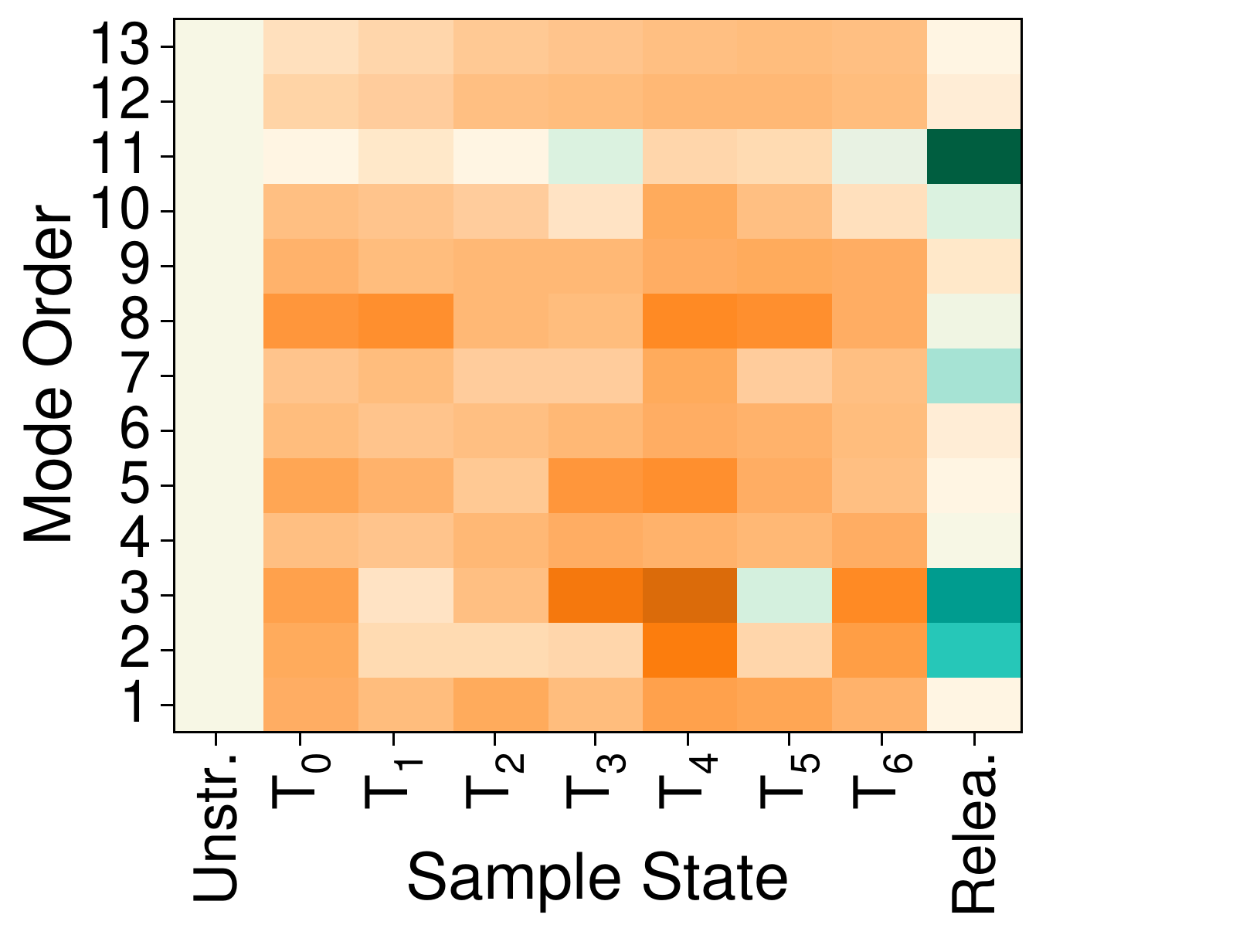}
    \end{subfigure}
    \vspace{-0.1cm} 

    \begin{subfigure}{0.43\textwidth}
        \centering
         \caption{}
        \includegraphics[width=\textwidth,trim=10 0 60 0,clip]{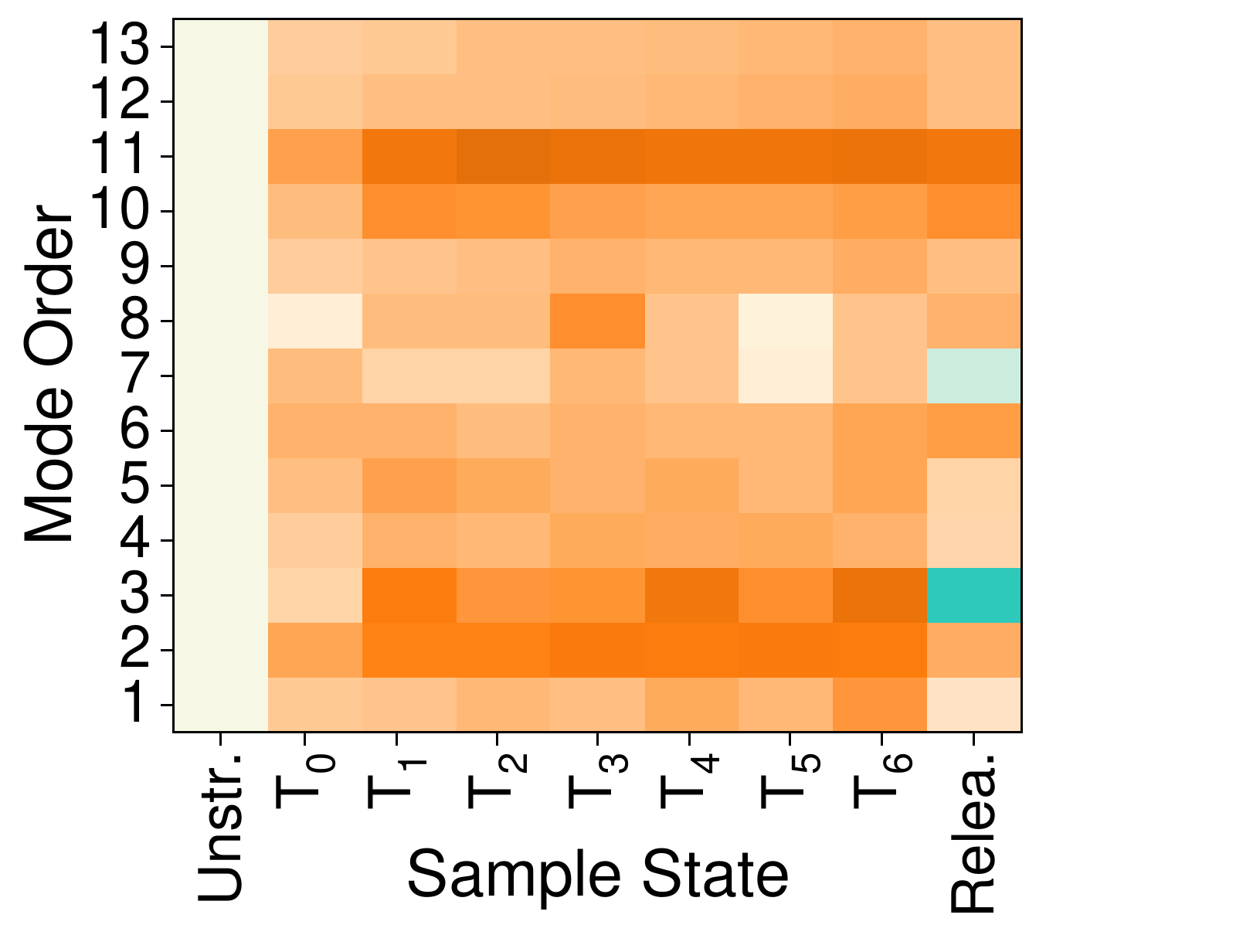}
    \end{subfigure}
    \hspace{0.02\textwidth} 
    \begin{subfigure}{0.43\textwidth}
        \centering
        \caption{}
        \includegraphics[width=\textwidth,trim=0 0 70 0,clip]{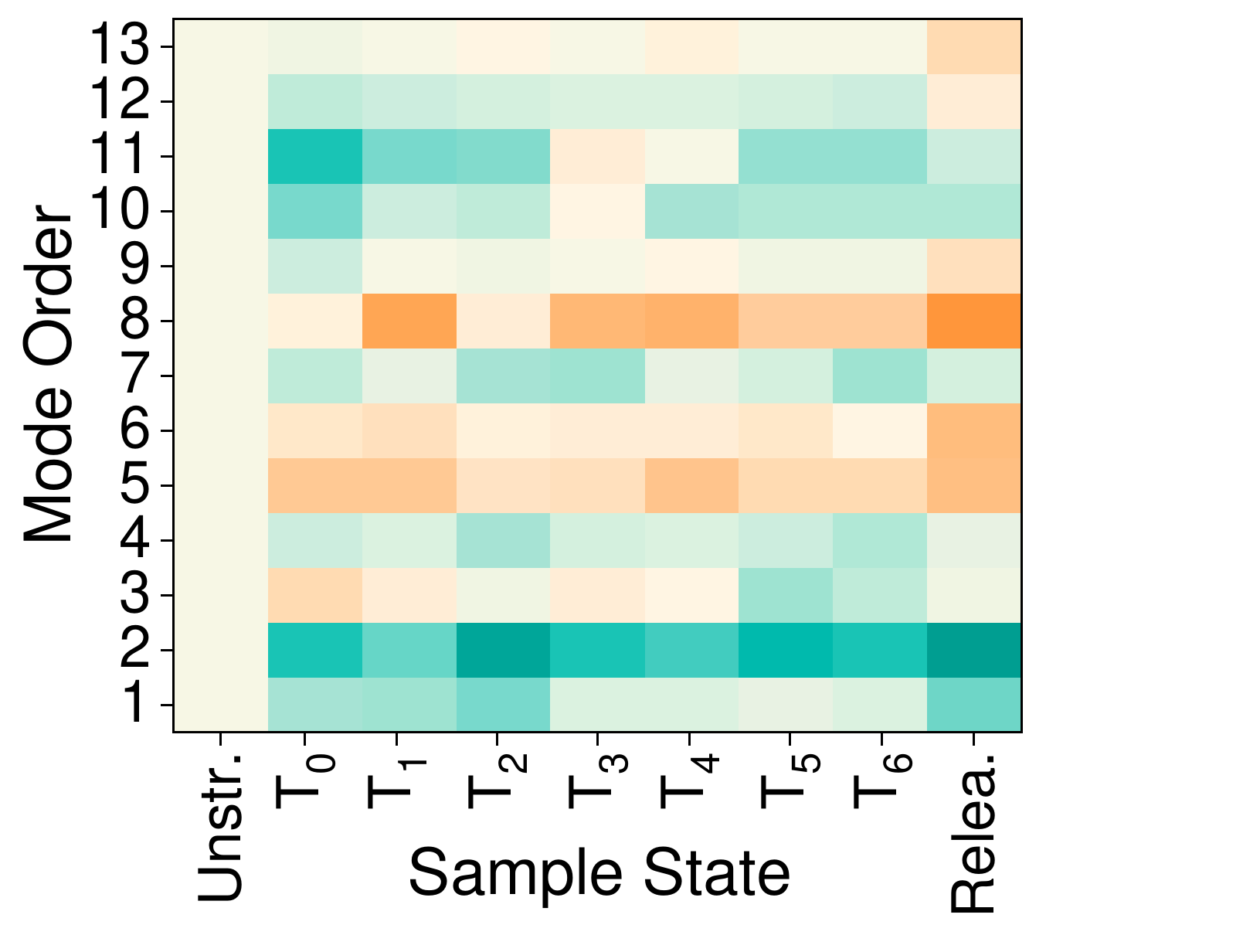}
    \end{subfigure}
    \vspace{-0.1cm} 

    \begin{subfigure}{0.43\textwidth}
        \centering
         \caption{}
        \includegraphics[width=\textwidth,trim=10 0 60 0,clip]{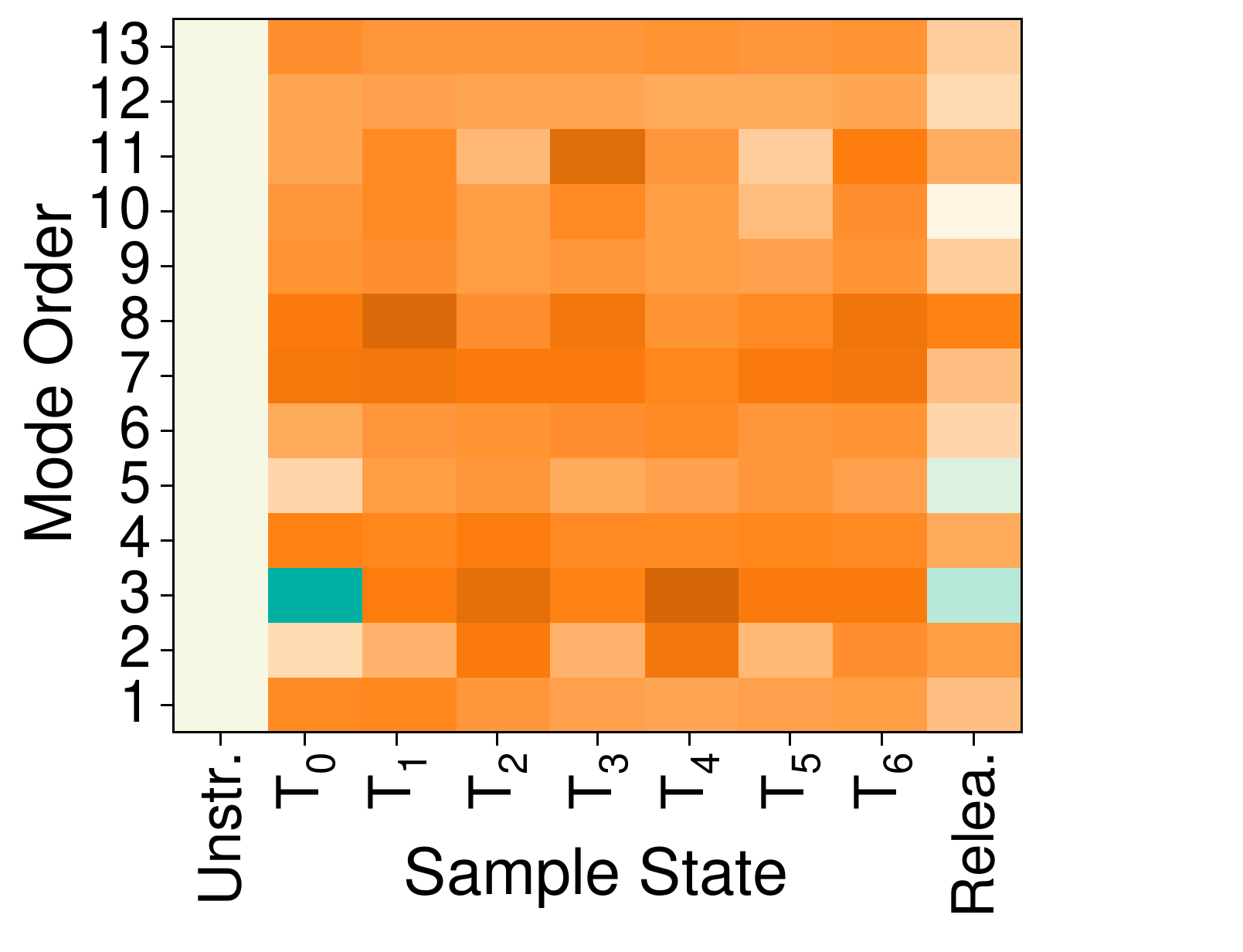}
    \end{subfigure}
    \hspace{0.02\textwidth} 
    \begin{subfigure}{0.43\textwidth}
        \centering
        \caption{}
        \includegraphics[width=\textwidth,trim=0 0 60 0,clip]{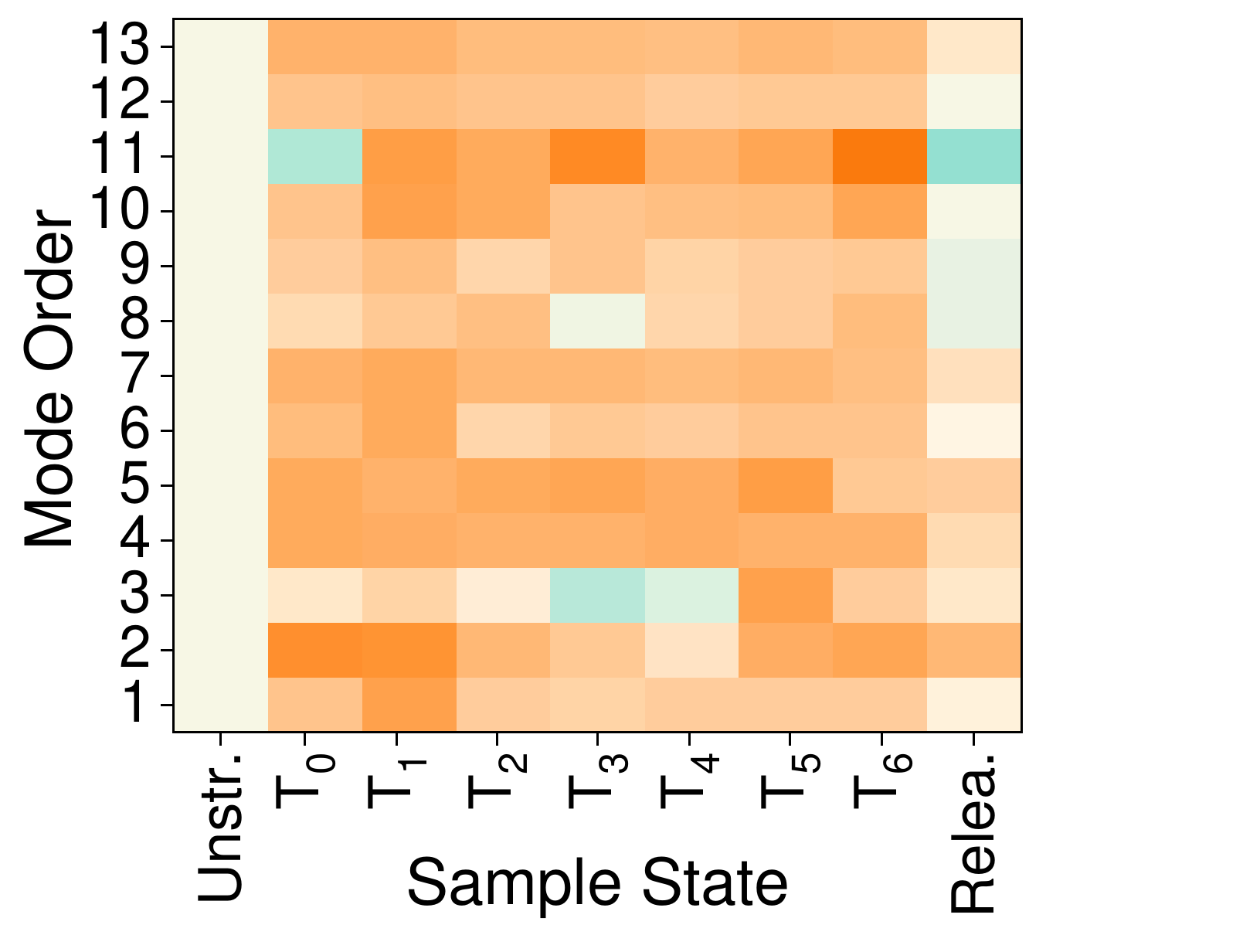}
    \end{subfigure}

    \caption{The graphs present the state evolution of normalized Raman line's intensity variations for PET under (a) 1\% (b) 5\%, (c) 10\%, (d) 15\%, (e) 20\%, and (f) 25\% applied tensile strain. The heatmaps compare the mode's intensity at various transitional states (x-axis) from $T_{0}$ to $T_{6}$ during strain application and after strain release (Relea.) The color bar is unified across different heatmaps. For each sample state, the measured value of each mode is normalized to the corresponding value in the unstretched state.}
    \label{chsoen strain-time intensity maps}
\end{figure}

Figure \ref{chsoen strain-time intensity maps} presents the intensity heatmaps for various levels of strain. Stretching by 1\% causes slight intensity variations initially, but an intense increase when the sample is exposed for a longer time, particularly after the initial 25 minutes (T\(_{5}\)) and after the release. The fifths and eleventh modes exhibit noticeable intensity changes (increased intensity) relative to the unstretched state. At 3\% strain, the intensity variations become more pronounced, and all the modes show strong intensity reduction from T\(_{3}\) to T\(_{6}\). Despite the wavenumber and FWHM, mostly returning to their initial values after strain release, the intensity does not fully recover to its original level.

For midrange strain values (5 to 10\%), strain application results in a decrease in the intensity of modes. 5\% strain introduces more intensity fluctuations across multiple modes. Notably, modes 3, 5, and 8 begin to exhibit stronger intensity reduction. The elastic behavior of the material at low strains likely redistributes the Raman active modes, leading to slight polarization-related intensity shifts. Most changes are reversible upon strain release, while certain modes including 2, 3, and 7 experience an increase in the line intensity. Increasing the strain to 10\% leads to a wide reduction in intensity, with modes 2, 3, and 11 showing a pronounced decrease. 

At 15\% strain, the overall intensity distribution becomes irregular. While the intensity distinctly increases in the low-wavenumber modes including modes 2 and 3, modes like 5, 6, and 8 experience a reduction. Limited reversibility is observed upon release. At 20-30\% strain, nearly all modes exhibit substantial intensity drop. At 30\% strain, the intensity heatmap shows a reduction in intensity for most modes, except the 5, which experiences a strong increase in all states. This suggests selective loss of Raman activity in certain vibration modes while the polarizability of mode 5 is increasing. The near-total loss of reversibility in FWHM and wavenumber data is consistent with the irreversible nature of intensity alterations. Strain-induced molecular alignment creates directional optical properties, which may benefit certain polarization-sensitive applications but compromise isotropic transparency. Table 4 presents a comprehensive overview of the observed mode intensity variations response.  \begin{table}[H]
\scriptsize
\centering
\begin{tabularx}{\textwidth}{|l|X|X|X|X|}
\hline
\rowcolor[gray]{0.8} 
\textbf{Strain Level} & \textbf{Key modes Affected} & \textbf{Intensity Patterns} & \textbf{Reversibility After Release} & \textbf{Conclusion} \\ \hline
\rowcolor[gray]{0.9} 
1\% & 5 & Minimal intensity changes for low wavenumbers, localized to specific modes & minimum reversibility & Early signs of molecular alignment, strain-induced changes begin\\ \hline
\rowcolor[gray]{0.95} 
2\% & 5, 11 & Small increases in intensity for sensitive modes & minimum reversibility & - \\ \hline
\rowcolor[gray]{0.9} 
3\% & All modes & high-intensity decrease with broader spread & Partial return to baseline & Increased molecular perturbations affecting polarizability. \\ \hline
\rowcolor[gray]{0.95} 
4\% & 5, 7, 11 & Noticeable intensity fluctuations across modes & Limited return & Enhanced alignment along the strain axis; tensor alterations visible. \\ \hline
\rowcolor[gray]{0.9} 
5\% & All modes & Intensity changes stabilize & partial residual changes recovery & Onset of permanent molecular rearrangement; altered polarizability. \\ \hline
\rowcolor[gray]{0.95} 
10\% & All modes & Significant intensity deviations & moderate residual effects & Irreversible alignment and structural deformation dominate the response. \\ \hline
\rowcolor[gray]{0.9} 
15\% & 2, 5, 8 & Broad intensity changes, concentrated in sensitive modes & Very limited recovery & Major structural reorganization impacts polarizability tensor. \\ \hline
\rowcolor[gray]{0.95} 
20\% & All modes & Intense deviations with permanent changes across modes & No recovery & Severe deformation, permanent polarizability alterations evident. \\ \hline
\rowcolor[gray]{0.9} 
25\% & 6th, 10th, 11th & Strong, consistent intensity deviations across spectrum & Very limited recovery & Irreversible structural reorganization and strong defect formation. \\ \hline
\rowcolor[gray]{0.95} 
30\% & All modes & Maximum intensity changes, broad and irreversible deviations & No recovery & Total loss of elasticity, complete reconfiguration of polarizability. \\ \hline
\end{tabularx}
\caption{Summary of strain-level effects and their impact on Raman mode intensity variation}
\label{raman intensity heatmaps table}
\end{table}

\subsubsection{Analysis of Gauche, Trans, and Amorphous Conformations Under Strain in PET}

\noindent \textbf{3.2.4.1 Modes Associated with Gauche and Trans Conformations} 

\noindent The 1119 cm$^{-1}$ band (mode 7), related to the gauche conformation of ethylene glycol (EG) units, exhibits alternating red and blue shifts, with the blue shift dominating for most strain values within the examined range. These shifts reflect the heterogeneous response of PET, where localized chain alignment restricts the rotational freedom of gauche conformers, which is indicative of prototypical tetragonal (proto-TX) phase formation. The proto-TX phase, as described in the literature, represents an intermediate structural state between the amorphous and crystalline phases, characterized by the dominance of trans conformers and localized molecular alignment \cite{cole}.
Furthermore, in the context of PET’s response to strain, heterogeneity refers to the localized and non-uniform molecular behavior within the material. For mode 7, the observed redshifts are more pronounced at higher strain values, suggesting a slight disruption in intermolecular packing. The FWHM decrease of this mode at low strains is another indicator consistent with the partial alignment of molecular chains.

The 998 cm$^{-1}$ band (mode 5), associated with the trans C–C bond, predominantly exhibits redshifts across all strain levels. At low strains, a pronounced red shift indicates a weakening of intermolecular interactions and packing density. Furthermore, the broadening of FWHM in this mode reflects localized strain-induced distortions.

The 1095 cm$^{-1}$ band (mode 6), associated with C–O stretching in trans conformers, displays smaller but consistent redshifts across all strain levels. This mode represents structurally stable regions, with its FWHM and intensity varying minimally compared to other modes, suggesting that C–O bonds are more resistant to strain-induced disruptions. However, the slight broadening of FWHM remains consistent with the previously mentioned strain-induced distortions.
\\ \textbf{3.2.4.2 Modes Associated with Amorphous and PET Structural Integrity}
\\The 1186 cm$^{-1}$ band (mode 8), attributed to CH$_{2}$ twisting and associated with amorphous PET, exhibits distinct behavior under strain. At low strain levels (1–5\%), its intensity fluctuates slightly, alternating between increases and decreases relative to the initial value, reflecting localized amorphous regions with a heterogeneous response. This behavior aligns with the proto-TX phase, where some amorphous regions persist while localized alignment begins. At higher strain levels ($>$5\%), the intensity consistently decreases, indicating a progressive loss of amorphous character as chain alignment and crystallinity increase. Additionally, the FWHM of mode 8 decreases across all strain levels, further reflecting the gradual transition from amorphous disorder to localized molecular alignment and more ordered states. Regarding the wavenumber shifts, alternating red and blue shifts are observed at low strains (1–5\%), serving as another indicator of the heterogeneous response of amorphous regions. Beyond 5\% strain, pronounced red shifts dominate, signaling significant packing disruptions and molecular bond elongation.

The 1725 cm$^{-1}$ band associated with the carbonyl (C=O) stretching mode is sensitive to the degree of crystallinity in PET, with higher frequencies and sharper peaks indicating a more crystalline structure.  At low strain levels (1–5\%), the band exhibits minor variations responding to strain. As the strain increases beyond 5\%, a redshift becomes more pronounced, indicating a significant weakening of the intermolecular interactions and an increased disruption of the density of the packing of chains. The reduction in the intensity of this band at higher strain levels aligns with growing disorder within the molecular structure, consistent with transitions observed in other vibrational modes.

\section{Correlation of Raman and Absorption Spectroscopy Findings}
The optical properties of PET under strain are intrinsically linked to its molecular structure, particularly the interplay between its amorphous and crystalline states, as well as strain-induced perturbations in molecular conformations and functional groups. In the amorphous state, PET consists of disordered polymer chains with a uniform refractive index, minimizing light scattering and maintaining high transparency. In contrast, the crystalline state introduces highly ordered and densely packed polymer chains, forming crystalline domains with refractive indices distinct from the surrounding amorphous regions. These boundaries scatter light, progressively reducing transparency as the degree of crystallinity increases. Strain exacerbates this effect by inducing structural transitions, creating defects, and perturbing chain packing, further enhancing light scattering and reducing transparency \cite{ward}.

The integration of Raman and UV-Vis findings underscores the critical role of strain-induced molecular alignment and phase transitions in governing PET's optical behavior. The detailed behavior of key vibrational modes—the 1119 cm$^{-1}$ (mode 7), 998 cm$^{-1}$ (mode 5), 1095 cm$^{-1}$ (mode 6), 1186 cm$^{-1}$ (mode 8), and 1725 cm$^{-1}$ (mode 13) bands—provides a molecular-level understanding of how strain influences structural evolution.

At low strain levels (1–4\%), PET maintains its transparency with negligible changes in optical properties. The Raman spectra suggest minimal molecular reorientation and negligible phase transitions, aligning with the absence of significant variations (up to 20\%) in the UV-Vis absorption spectra, indicating preserved transparency and molecular integrity. 

As strain increases to intermediate levels (5–10\%), PET’s optical properties begin to show measurable changes. The Raman study offers a unique perspective on molecular-level order sensitivity in the response of PET to mechanical strain. This paves the way for future research to explore polarization-dependent measurements and the engineering of PET films with controlled molecular orientations, which introduce variations in the refractive index and result in light scattering and reduced transparency. UV-Vis spectra corroborate these findings, showing an increase in absorbance to up to 90\% for the 10\% applied strain in the visible spectrum, particularly in the blue and purple regions, due to structural transitions. 

At higher strain levels (15–30\%), PET undergoes irreversible structural changes, as revealed by Raman spectroscopy. These changes include significant peak broadening and intensity reduction in critical vibrational modes, indicating the breakdown of molecular order and extensive chain disruption. UV-Vis absorption spectroscopy shows a more than 100\% increase in absorbance, which corresponds to enhanced scattering due to the formation of crystalline domains and a reduction in the optical band gap, likely caused by an increase in defect density within the material. These structural changes lead to severe transparency loss and degradation of optical clarity, driven by residual stresses and increased structural non-uniformity. Additionally, optical anisotropy becomes pronounced, as strain-induced molecular alignment creates directionally dependent optical properties. Raman intensity variations further emphasize PET’s sensitivity to strain, with an average 30\% reduction in Raman line intensity across all modes at 20\% applied strain.

As mentioned earlier, the Raman spectroscopy measurements were conducted while the PET samples were held under strain for 30 minutes at room temperature. Instead of a single transient state, multiple transient states were observed in the intermediate spectra, reflecting ongoing molecular adjustments. This behavior can be attributed to the inherent rigidity of PET at room temperature, which influences molecular mobility and limits immediate large-scale rearrangements. The observations suggest that PET’s molecular response under strain at room temperature primarily involves localized and gradual adjustments, rather than long-range relaxations occurring within a short time frame. This structural rigidity may also explain the dominance of heterogeneous responses in key Raman modes, as strain-induced changes in the amorphous and crystalline regions remain localized.

\subsection{Practical Implications of PET’s Strain Response}

Manufacturers can develop PET films with improved strain resistance, ensuring reliability in demanding environments by understanding the strain thresholds at which irreversible changes occur (e.g., above 10–15\% strain). The pronounced intensity, FWHM, and wavenumber variations in Raman spectra under strain could be harnessed to create susceptible, polarization-dependent Raman sensors \cite{Roisin2023}. These sensors would be capable of precisely detecting localized stress or strain, making them ideal for applications in structural health monitoring or wearable electronics.
By correlating strain-induced structural changes with optical degradation, the study offers a framework for predicting the operational stability of PET-based devices, allowing for better material selection and design \cite{Lipomi2012}.

Moreover, the implications of this work can extend to other industries where PET is widely used, such as packaging and structural films. In mechanical engineering, PET’s robustness under strain is crucial for applications like protective films and load-bearing components. The insights from this study can guide the development of PET materials with enhanced mechanical properties, such as higher tensile strength and improved elasticity. Understanding how strain impacts PET at a molecular level can aid in designing packaging materials that maintain transparency and durability under various stresses, contributing to sustainable and high-performance solutions. 

The study’s focus on fixed polarization offers a unique perspective on the anisotropic response of PET to mechanical strain. This could pave the way for future research exploring developing Raman sensors optimized for different polarization configurations to enhance strain detection. As well as engineering PET films with controlled molecular orientations to achieve specific optical and mechanical characteristics for niche applications.
\section {Conclusion}
This study provides a comprehensive understanding of the optical and structural responses of PET films to mechanical strain at room temperature, offering insights distinct from those observed under elevated-temperature conditions. By combining UV-Vis absorption and Raman spectroscopy, this work reveals the intricate interplay between molecular structure and optical behavior, uncovering strain-induced phenomena that define PET’s performance under mechanical deformation.

At strain levels below 5\%, PET retains its transparency and molecular integrity, reflecting minimal molecular alignment and negligible crystallinity changes. These properties highlight its suitability for applications requiring small deformations, such as optical coatings and flexible sensors. Beyond this threshold, structural transitions become pronounced, with the emergence of molecular alignment and increased crystallinity inducing optical anisotropy and light scattering, ultimately reducing transparency. Such insights are critical for setting strain limits in applications like flexible displays and photovoltaic layers, where optical performance is paramount.

At higher strain levels (15–30\%), irreversible structural changes dominate. These include significant disruptions in chain packing, defect formation, and the collapse of isotropic optical properties. These changes result in a pronounced loss of transparency and elasticity but open pathways for utilizing PET in strain-sensitive applications, such as optical strain sensors and polarization-dependent devices.

This work underscores the unique insights gained from room-temperature investigations, where restricted molecular mobility provides a clearer picture of how PET behaves in real-world operating conditions. The correlation of optical and structural changes at various strain levels offers a holistic understanding that informs the design of PET-based materials with enhanced optical and mechanical performance. These findings provide a foundation for advancing flexible electronics, wearable devices, and next-generation optical systems where PET’s strain response is both a challenge and an opportunity.

\begin{acknowledgement}

\end{acknowledgement}

\begin{suppinfo}

\end{suppinfo}

\bibliography{achemso-demo}

\end{document}